\documentclass[aps,twocolumn,superscriptaddress]{revtex4-1}

\usepackage[latin1]{inputenc}
\usepackage{float}
\usepackage{amsmath}
\usepackage{mathbbol}
\usepackage{amsfonts}
\usepackage{graphicx}
\usepackage{color}
\usepackage[normalem]{ulem}

\DeclareSymbolFontAlphabet{\amsmathbb}{AMSb}%

\newcommand{\vect}[1]{\boldsymbol{#1}}

\begin{document}
\title{Double-layer force suppression between charged microspheres}
\date{\today}
%%---------------------------------------
\author{D. S. Ether}
\email[]{ether@if.ufrj.br}
\homepage[]{http://sites.if.ufrj.br/lpo/en/}
\affiliation{Instituto de F\'{\i}sica, Universidade Federal do Rio de Janeiro, Caixa Postal 68528, Rio de Janeiro, RJ 21941-972, Brazil}
\affiliation{Laborat\'orio de Pin\c{c}as \'Oticas - LPO-COPEA, Instituto de Ci\^encias Biom\'edicas, Universidade Federal do Rio de Janeiro, Rio de Janeiro, RJ 21941-902, Brazil}
%%---------------------------------------
\author{F. S. S. Rosa}
\affiliation{Instituto de F\'{\i}sica, Universidade Federal do Rio de
Janeiro, Caixa Postal 68528, Rio de Janeiro, RJ 21941-972, Brazil}
%%---------------------------------------
\author{D. M. Tibaduiza}
\affiliation{Instituto de F\'{\i}sica, Universidade Federal do Rio de
Janeiro, Caixa Postal 68528, Rio de Janeiro, RJ 21941-972, Brazil}
%%---------------------------------------
\author{L. B. Pires}
\affiliation{Instituto de F\'{\i}sica, Universidade Federal do Rio de
Janeiro, Caixa Postal 68528, Rio de Janeiro, RJ 21941-972, Brazil}
\affiliation{Laborat\'orio de Pin\c{c}as \'Oticas - LPO-COPEA, Instituto de Ci\^encias Biom\'edicas, Universidade Federal do Rio de Janeiro, Rio de Janeiro, RJ 21941-902, Brazil}
%%---------------------------------------
\author{R. S. Decca}
\affiliation{Department of Physics, Indiana University-Purdue University Indianapolis, Indianapolis, Indiana 46202, USA}
%%---------------------------------------
\author{P. A. Maia Neto}
\affiliation{Instituto de F\'{\i}sica, Universidade Federal do Rio de
Janeiro, Caixa Postal 68528, Rio de Janeiro, RJ 21941-972, Brazil}
\affiliation{Laborat\'orio de Pin\c{c}as \'Oticas - LPO-COPEA, Instituto de Ci\^encias Biom\'edicas, Universidade Federal do Rio de Janeiro, Rio de Janeiro, RJ 21941-902, Brazil}
%%---------------------------------------
\date{\today}
\begin{abstract}
\par In this paper we propose a protocol to suppress double-layer forces between two microspheres immersed in a dielectric medium, being one microsphere metallic at a controlled potential $\psi_{\rm{M}}$ and the other a charged one either metallic or dielectric. The approach is valid for a wide range of distances between them. We show that, for a given distance between the two microspheres, the double-layer force can be totally suppressed by simply tuning $\psi_{\rm{M}}$ up to values dictated by the linearized Poisson-Boltzmann equation. Our key finding is that such values can be substantially different from the ones predicted by the commonly used  proximity force approximation (PFA), also known as Derjaguin approximation, even in situations where the latter is expected to be accurate. The proposed procedure can be used to suppress the double-layer interaction in force spectroscopy experiments, thus paving the way for measurements of other surface interactions, such as Casimir dispersion forces.
\end{abstract}


\maketitle

\section{\label{sec:intro}Introduction}
\par Electrostatic double-layer forces are among the most important interactions between solid surfaces in liquids \cite{butt10,israelachvili11}. Being one of the pillars of the Derjaguin, Landau, Verwey, and Overbeek (DLVO) theory \cite{Verwey48, deryaguin39,deryaguin41}, which has been developed seven decades ago to explain the aggregation of aqueous dispersions, this interaction has been extensively studied over the last century, from the pioneering works of Gouy, Chapman, and Debye \cite{chapman13,gouy10,debye23} to the recent force spectroscopy experiments 
using atomic force microscopes (AFM)~\cite{butt05,munday07,munday08,zwol09,popa10,borkovec12,Wodka14}, optical tweezers (OT)~\cite{crocker94,hansen05,sainis07,sainis08,schaffer07,ether15}, 
and total internal reflection microscopy (TIRM)~\cite{flicker93,liu14,liu16}. 
\par In some experiments, however, electrostatic double-layer forces were actually suppressed in order to probe other interactions such as the Casimir force \cite{munday09}. In fact, electrostatic double-layer force suppression is commonly achieved by salt screening of the surface charges in polar media, \cite{munday09}, or by using specific soluble charge control agents in apolar environments \cite{sainis08}. When metallic surfaces are present, another strategy consists in finding particular values of the electrostatic potentials at those surfaces that suppress the double-layer force between them. These values, henceforth named potentials of zero force (PZF)~\cite{hillier96}, are obtained by either changing the solution's pH or by electrostatic potential tuning~\cite{hillier96,barten03,munday08}. 
% check this....
\par double-layer forces are usually described by the (non-linear) Poisson-Boltzmann (PB) equation \cite{butt05}. 
When the surfaces are separated by very short distances (a few nanometers or less), two important properties determine the nature of the interaction:
(i) the potentials involved are typically much larger than $k_B T/e$ (where $e$ is the elementary charge), so that non-linearities are relevant; and (ii) the curvature of the surfaces has a negligible effect, so one can use the proximity force approximation (PFA) to calculate the force by averaging the result for parallel planar surfaces over the local distance.
 In this work, however, we are interested in larger distances (tens of nanometers or more),
 which is, for instance,  the relevant range for Casimir force experiments. 
In this case, we are entitled to replace the full PB equation by its  linearized version -- henceforth referred to as Debye-H\"{u}ckel  (DH) equation  \cite{butt05}. On the other hand, the curvature effects become more relevant, and it is indispensable to take the spherical geometry exactly into account. 
\par Henceforth, we propose a theoretical protocol to suppress the double-layer force between two charged microspheres immersed in a dielectric medium for a wide distance range. 
One of the microspheres is metallic and placed at an externally controlled electrostatic potential $\psi_{\rm{M}}$, while the other microsphere could be either (i) a charge regulated (CR) dielectric microsphere, or (ii) an electrically isolated metallic microsphere with total charge $Q$. Then, for a given distance $L$, the electrostatic double-layer force between them can be totally suppressed by simply tuning $\psi_{\rm{M}}$ up to a calculated PZF value. Our most striking observation is that there are situations where the PFA predictions for the PZFs are completely off the mark, even when they are, in principle, expected to be accurate. As we shall discuss, this is not exactly due to a failure of the PFA, but rather to the fact that the PFA is not well defined in such situations.  

The paper is organized as follows. The results for dielectric and metallic spheres are presented in Secs. II and III, respectively. Sec.~IV is dedicated to concluding remarks. Appendices A and B contain details of the theoretical derivation, whereas appendices C and D discuss commonly employed approximations. 

%obtained by solving a second order polynomial equation, with coefficients related to the (exact) solutions of the DH equation \cite{glendinning83, carnie93,carnie94}

%This method does not rely on pH modifications of the solutions nor salt screening, and gives PZF values that, depending on the situation, can considerably differ from the values calculated by LSA and PFA approximations. 

 %For the case of two microspheres, force expressions in the DH regime have been obtained in terms of multipole expansions \cite{glendinning83, carnie93,carnie94}, although 

%However, in many of the related experiments \cite{crocker94,hansen05,sainis07,sainis08,schaffer07,gutsche07,zwol09,popa10,borkovec12,Wodka14}, due to electrostatic screening and/or the distance range, only approximate expressions were needed, such as the linear superposition approximation (LSA) or proximity force approximation (PFA). In fact, in \cite{ether15}, data suggest deviations from LSA as the distance between microspheres decreases.

%
\section{ \label{sec:dielreg} Charged regulated dielectric microsphere}
\subsection{\label{sec:model} Debye-H\"uckel theory with two spherical surfaces}
\par
\begin{figure}
\centering
\hspace{-10pt}
\includegraphics[scale=0.30]{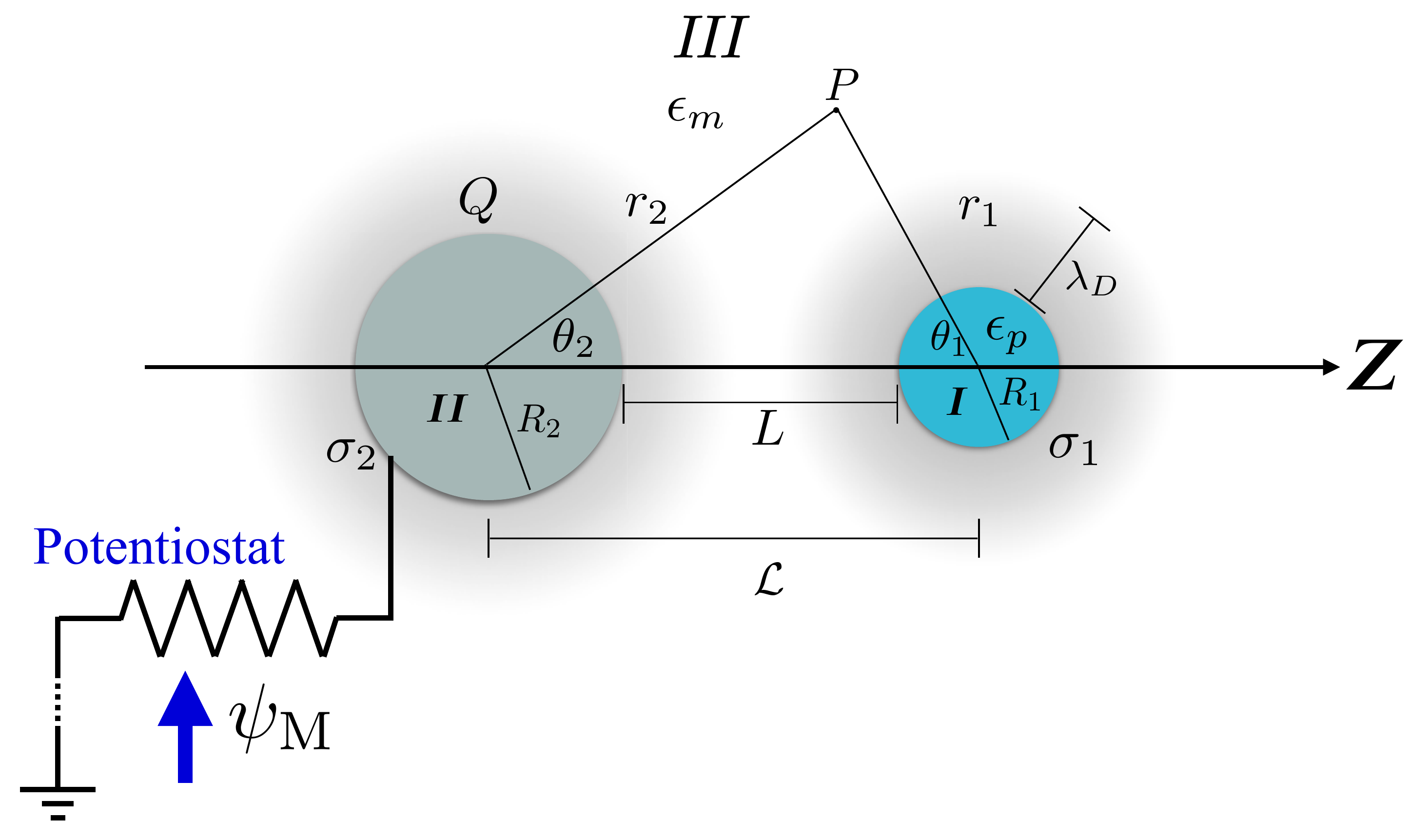}
\caption{(Color online) Dielectric and metallic microspheres.  The potentiostat controls and fixed the metallic microsphere electrical potential $\psi_{\rm{M}}$.}
\label{fig:1}
\end{figure}

 We consider a linear and isotropic dielectric microsphere of radius $R_{1}$, relative permittivity $\epsilon_{p}$, and a metallic microsphere of radius $R_{2}$, with their centers separated by a distance $\mathcal{L}$ along the $z$ axis,  as depicted in Fig.\ref{fig:1}. The distance of closest approach is $L=\mathcal{L}-R_1-R_2$. The microspheres are embedded in an isotropic $Z:Z$ electrolyte, which is a substance which separates into cations and anions of $Z$ and $-Z$ valence under solvation, with relative permittivity $\epsilon_{m}$, and the whole system is in thermal equilibrium at a temperature $T$. We assume that the metallic sphere is held at an externally controlled electrostatic potential $\psi_{\rm{M}}$, and that the dielectric microsphere may exchange charge with the medium by ion adsorption or dissociation processes \cite{trefalt16}.
\par
The electrostatic potential in the electrolyte outside the microspheres (region $I\!I\!I$ in Fig.~\ref{fig:1}) satisfies the DH equation~\cite{debye23,butt10, israelachvili11}
\begin{equation}
\left(\nabla^{2}-\kappa^{2}\right)\psi(\mathbf{r})=0\,,
\label{eq:lpbt}
\end{equation}
where
\begin{equation}
\lambda_{\rm D}=\frac{1}{\kappa}= \sqrt{\frac{\epsilon_m k_{B}T}{2(Ze)^{2} n_{\infty}}}
\label{eq:DebyeL}
\end{equation}
is the Debye length, a measure of the diffuse double-layer thickness \cite{butt10}. $n_{\infty}$ is the ionic bulk concentration (assumed to be much higher than the concentration of ions dissociated from the surfaces).
While the linear approximation leading to the DH equation (\ref{eq:lpbt}) is strictly valid when $\left|\psi\right|\lesssim25\,{\rm mV}$ for  $T\sim300\,{\rm K}$, it is commonly found for the results to be trustworthy for  potentials up to $50-80\,\rm{mV}$ \cite{butt10}. 
\par 
We assume that there is no free electric charge inside the microspheres. Thus, the electrostatic potential in regions  $I$ and $I\!I$ satisfies the Laplace's equation
\begin{equation}
\nabla^{2}\psi(\mathbf{r})=0\,.
\label{eq:lap}
\end{equation}
\subsubsection{General solution outside and inside the microspheres}
\par Following \cite{glendinning83,carnie93,carnie94}, we  write the general solution of  eq.~(\ref{eq:lpbt}) with azimuthal symmetry 
as
\begin{eqnarray}
\psi^{I\!I\!I}(P)&=&\sum_{n=0}^{\infty}\left[a_{n}k_{n}(\kappa r_{1}) 
P_{n}(\cos \theta_1) \right. \nonumber \\
&& \left. \hspace{15pt }+ \, b_{n}k_{n}(\kappa r_{2})P_{n}(\cos 
\theta_2)\right]\,,
\label{eq:psiout12t}
\end{eqnarray}
where
$r_1$ and $\theta_1$ are the spherical coordinates of a point $P$  in region $I\!I\!I$ with respect to the center of sphere $I,$ and likewise for $r_2$ and $\theta_2$ with respect 
to the center of  sphere $II,$ as illustrated in Fig.~\ref{fig:1}. $P_{n}$ is the $n$-th order Legendre polynomial and $k_{n}$ is the modified spherical Bessel function of the third kind and order $n$ defined in appendix \ref{ap:A}. The  unknown coefficients $a_{n}$ and $b_{n}$ will be found by applying appropriate boundary conditions. 
 \par In region $I$, the azimuthal symmetrical solution to Eq. (\ref{eq:lap}) is given by \cite{jackson99}
\begin{eqnarray}
\psi^{I}(r_{1},\theta_{1})&=& \sum_{n=0}^{\infty} c_{n}r_{1}^{n} P_ {n} (\cos \theta_{1})\,,
\label{eq:psiin1t}
\end{eqnarray} 
with $c_{n}$ the corresponding unknown coefficients. In  region $I\!I,$ the electrostatic potential is fixed:
\begin{eqnarray}
\psi^{I\!I}(r_{2},\theta_{2})&=&\psi_{\rm{M}}\,.
\label{eq:psiin2t}
\end{eqnarray} 
\subsubsection{\label{sec:Bcond}Boundary Conditions}
\par
The boundary conditions for the electrostatic potential at the surface of the dielectric microsphere are given by 
\begin{widetext}
\begin{eqnarray}
\left.\psi^{I}(r_{1},\theta_{1})\right|_{r_1=R_1^{-}}=\left.\psi^{I\!I\!I}(r_{1},\theta_{1})\right|_{r_1=R_1^{+}}\,,\label{eq:bcond1}&&\\
\vect{\hat{n}}\cdot\left[\left.\epsilon_{p}\nabla\psi^{I}(r_{1},\theta_{1})\right|_{r_1=R_1^{-}}-\left.\epsilon_{m}\nabla\psi^{I\!I\!I}(r_{1},\theta_{1})\right|_{r_1=R_1^{+}}\right]&=&\frac{\sigma_{1}\left(\theta_{1}; L\right)}{\epsilon_{0}}\,,
\label{eq:bcond2}
\end{eqnarray}
\end{widetext}
where $\epsilon_0$ is the vacuum permittivity,  $\vect{\hat{n}}$ is the outward normal unit vector and $\sigma_{1}\left(\theta_{1}; L\right)$ is the non-uniform surface charge density  that in general depends on the separation $L.$ 
We consider a constant charge regulation model (CR)~\cite{krozel91,carnie93,behrens99a,behrens99b,pericet-camara04,ruiz-cabello13,trefalt16}:
\begin{equation}
\sigma_{1}\left(\theta_{1}; L\right)=\sigma_{01}-{\rm C}^{\rm{sph}}_{I}\left.[\psi^{I\!I\!I}(R_1^{+},\theta_{1}; L)-\psi_{01}\right]\,,
\label{eq:CR}
\end{equation}
where $\sigma_{01}$ and $\psi_{01}$ are respectively the surface charge density and potential of the dielectric microsphere when 
 isolated, \emph{i.e.}, when $ L\gg \lambda_{\rm D}$. They are related by 
the capacitance per unit area ${\rm C}^{\rm{sph}}_{D}$ of the isolated dielectric microsphere:
\begin{eqnarray}
\psi_{01} &= &\frac{\sigma_{01}}{{\rm C}^{\rm{sph}}_{D}} \label{psi01a}\\ 
{\rm C}^{\rm{sph}}_{D}&=&\frac{\epsilon_0\, \epsilon_m\left(1+\kappa R_{1}\right)}{R_{1}}.\label{psi01b}
\end{eqnarray}
The constant ${\rm C}^{\rm{sph}}_{I}=-\partial\sigma_{1}/\partial\psi\geq0$ is the regulation capacitance per unit area.
  It quantifies the dielectric surface dissociation rates and
  is assumed to be distance independent. 
  In the DH regime, the Boltzmann factor is linearized, and then the potential difference appearing in the r.-h.-s. of (\ref{eq:CR}) represents the 
  electrolyte charge density variation with distance in the vicinity of the sphere surface. Thus, the CR model (\ref{eq:CR}) yields the local surface charge density
  $\sigma_1$
   as a linear response to the modification of the 
  ionic concentration in the vicinity of the surface. The resulting distribution is invariant under rotations around the $z$ axis, hence leading to a potential with azimuthal symmetry as given by
Eqs.~(\ref{eq:psiout12t}) and (\ref{eq:psiin1t}).
  
  Two limiting cases are noteworthy:
  when ${\rm C}^{\rm{sph}}_{I}\rightarrow 0$, we recover the constant charge (CC) model,
  in which the surface charge density is assumed to be uniform, distance independent and prescribed. 
In the opposite limit ${\rm C}^{\rm{sph}}_{I}\rightarrow\infty,$ we have the constant potential (CP) model. It is convenient to define the  regulation parameter $p$ \cite{trefalt13,pericet-camara04,trefalt16}:
\begin{equation}
p=\frac{{\rm C}^{\rm{sph}}_{D}}{{\rm C}^{\rm{sph}}_{D}+{\rm C}^{\rm{sph}}_{I}}\,.
\label{eq:psphere}
\end{equation}
The CP and CC limits correspond  to $p=0$ and $p=1,$ respectively. Intermediate values correspond to a situation in which both charge density and potential of the dielectric sphere are perturbed as the metallic sphere approaches. 
\par For the metallic microsphere, we have the boundary conditions 
\begin{eqnarray}
\left.\psi^{I\!I\!I}(r_{2},\theta_{2})\right|_{r_2=R_2^{+}}&=&\psi_{\rm{M}}\label{eq:bcondmetal1}\,,\\
-\left.\epsilon_{m}\vect{\hat{n}}\cdot\nabla\psi^{I\!I\!I}(r_{2},\theta_{2})\right|_{r_2=R_2^{+}}&=&\frac{\sigma_{2}\left(\theta_{2}; L\right)}{\epsilon_{0}}\,,
\label{eq:bcondmetal2}
\end{eqnarray}
where again $\vect{\hat{n}}$ is the outward normal unit vector and $\sigma_{2}\left(\theta_{2}; L\right)$ is the non-uniform metallic surface charge density. 
\subsubsection{Electrostatic potential in the electrolyte}
\par 
We solve for the coefficients $a_n$ and $b_n$ giving the potential in the electrolyte [see Eq.~(\ref{eq:psiout12t})] in terms of the metallic sphere
prescribed potential $\psi_{\rm{M}}$ and the  dielectric sphere unperturbed surface charge density $\sigma_{01}$ (\emph{i.e.}, the charge density for $L\gg 1/
\kappa$). 
We first define 
\begin{eqnarray}
X_{n}\equiv A_{n}a_{n}& \hspace{10pt} , \hspace{10pt} Y_{n}\equiv k_{n}(\kappa R_2) b_{n}\,,
\label{eq:XYsyst}
\end{eqnarray}
where 
\begin{equation}
A_{n}\equiv\left[n\,\frac{\epsilon_{p}}{\epsilon_{m}}+\frac{R_{1}{\rm C}^{\rm{sph}}_{I}}{\epsilon_0\, \epsilon_m}\right] k_{n}(\kappa R_{1})-\kappa R_{1}k_{n}^{\prime}(\kappa R_{1}),
\label{eq:Acoeff}
\end{equation}
with $k_{n}^{\prime}$ the derivative of the modified spherical Bessel function of the third kind $k_{n}$ with respect to its argument. We now match the expressions for the potential in the three regions, as given by Eqs.~(\ref{eq:psiout12t}), (\ref{eq:psiin1t}) and (\ref{eq:psiin2t}), taking the boundary conditions (\ref{eq:bcond1}), (\ref{eq:bcond2}), and (\ref{eq:bcondmetal1}) into account. We use the addition  theorem for Bessel functions in order to express the potential in the electrolyte, given by Eq.~(\ref{eq:psiout12t}), in terms of a single coordinate system, as detailed in Appendix A. We also replace the dielectric charge density appearing in  (\ref{eq:bcond2}) by the CR model Eq.~(\ref{eq:CR}). The resulting system of coupled linear equations is written as
\begin{eqnarray}
&&\mathbf{X} + \amsmathbb{B} \cdot \mathbf{Y} = \frac{R_{1}\bar{\sigma}_{1}}{\epsilon_0\, \epsilon_m} \mathbf{e}_0\,, \nonumber \\
&&\mathbf{Y} + \amsmathbb{C} \cdot \mathbf{X} = \psi_{\rm{M}} \, \mathbf{e}_0\,, 
\label{eq:lsysXYt}
\end{eqnarray}
where $e_{0i}= \delta_{i0}$,
\begin{eqnarray}
\bar{\sigma}_{1}&\equiv & \sigma_{01}+{\rm C}^{\rm{sph}}_{I}\psi_{01}=\frac{\sigma_{01}}{p}\,, \\
\label{eq:sigmabar}
\amsmathbb{B}_{jk}&\equiv&(2j+1)\frac{B_{jk}(\kappa \mathcal{L})}{k_k(\kappa R_2)}\left\{\left[j\, \frac{\epsilon_{p}}{\epsilon_{m}}+\frac{R_{1}{\rm C}^{\rm{sph}}_{I}}{\epsilon_0\, \epsilon_m}\right] i_{j}(\kappa R_{1})\right.\nonumber\\
&&\left.-\kappa R_{1}i_{j}^{\prime}(\kappa R_{1})\right\}\,,\label{eq:Bcoeff}\\
\amsmathbb{C}_{jk}&\equiv&(2j+1)\frac{B_{jk}(\kappa \mathcal{L})}{A_k}i_{j}(\kappa R_{2})\,.
\label{eq:Ccoeff}
\end{eqnarray}
We have again used the prime to denote the derivative of the modified spherical Bessel function of the first kind  $i_{j}$ (defined in appendix \ref{ap:A}) with respect to its argument. 
$B_{jk}(\kappa \mathcal{L})$ is defined by Eq.~(\ref{eq:Bcoef}) in Appendix \ref{ap:A}.
\par We solve the system (\ref{eq:lsysXYt}) and find
\begin{eqnarray}
X_{n}&=&\frac{R_{1}\bar{\sigma}_{1}}{\epsilon_0\, \epsilon_m}(\amsmathbb{E}^{-1})_{n0}-\psi_{\rm{M}}\amsmathbb{F}_{n0}\,,\\
Y_{n}&=&\psi_{\rm{M}}\amsmathbb{G}_{n0}-\frac{R_{1}\bar{\sigma}_{1}}{\epsilon_0\, \epsilon_m}\amsmathbb{H}_{n0}\,,
\label{eq:akpsicondarb}
\end{eqnarray}
where
\begin{eqnarray}
\amsmathbb{E} &\equiv& \mathbb{1} - \amsmathbb{B} \cdot \amsmathbb{C}\label{eq:Esys}\,,\\
\amsmathbb{F} &\equiv& \amsmathbb{E}^{-1} \cdot \amsmathbb{B}\label{eq:Fsys}\,,\\
\amsmathbb{G}&\equiv& \mathbb{1} + \amsmathbb{C} \cdot \amsmathbb{F}\label{eq:Gsys}\,,\\
\amsmathbb{H} &\equiv&  \amsmathbb{C} \cdot \amsmathbb{E}^{-1}\label{eq:Hsys}\,,
\end{eqnarray}
and with $\mathbb{1}$ denoting the identity matrix.

 Once the linear system Eq.(\ref{eq:lsysXYt}) is solved, we obtain the coefficients $a_{n}$ and $b_{n}$ from  Eq.~(\ref{eq:XYsyst}).
 The potential in the electrolyte is then known from 
 Eq.~(\ref{eq:psiout12t}),  allowing us to calculate the double-layer force between the microspheres as described below.
\subsection{\label{sec:quadforce} DH double-layer force and potential of zero force (PZF)}
\par The double-layer force on the dielectric microsphere is given by \cite{krozel91,carnie93,carnie94,glendinning83}
\begin{equation}
\vect{\rm{F}}_{1}\equiv\oint_{\partial\Re_{1}} \overleftrightarrow{\vect{\Theta}}\cdot\vect{\hat{n}}\,dA\,,
\label{eq:F1}
\end{equation}
where $\partial\Re_{1}$ is a surface  enclosing the sphere and
\begin{equation}
\overleftrightarrow{\vect{\Theta}}\equiv\overleftrightarrow{\vect{T}}-\Pi\overleftrightarrow{\vect{1}}\,,
\label{eq:thetatensorM}
\end{equation}
is the tensor accounting for the pressures and stresses in the electrolyte. The latter contains an electrical contribution given by 
the Maxwell stress tensor
\begin{equation}
T_{ij}=\epsilon_0\, \epsilon_m\left(E_{i}E_{j}-\frac{1}{2}\delta_{ij}E^{2}\right)\,
\label{eq:maxwellstresstensor}
\end{equation} 
  and the osmotic pressure contribution
\begin{equation}
\Pi\equiv\frac{\epsilon_0\, \epsilon_m\kappa^{2}\psi^{2}}{2}
\label{eq:osmotic}
\end{equation}
arising from the non-uniform  ionic volume density. 
\par 
Choosing $\partial\Re_{1}$ to be a spherical surface of radius $R_{1}+\delta\,(\mbox{with}\, \delta\rightarrow 0^+)$, using the obtained $a_{n}$ and $b_{n}$ coefficients in (\ref{eq:psiout12t}) to calculate all the electric field components in Eqs.~(\ref{eq:maxwellstresstensor}), and remembering that azimuthal symmetry entails that the force is along the $z$ axis, the double-layer force is given by 
\begin{equation}
\rm{F}_{\rm{D}}(\mathit{L})=\mathcal{A}_{\rm {D}}(\mathit{L})\psi^{2}_{M}+\mathcal{B}_{\rm {D}}(\mathit{L})\psi_{\rm{M}}+\mathcal{C}_{\rm {D}}(\mathit{L})\,,
\label{eq:DLMDt}
\end{equation}
which has a simple quadratic dependence on the metallic sphere potential $\psi_{\rm{M}}$. The  coefficients $\mathcal{A}_{\rm D}\left(L\right)$,  $\mathcal{B}_{\rm D}\left(L\right)$ and  $\mathcal{C}_{\rm D}\left(L\right)$ are fairly complicated functions of $L$, and explicit expressions are given in Appendix \ref{ap:B}. 
 
 Eq.~(\ref{eq:DLMDt})  tells us  how to produce a vanishing force for a given distance $L$: one has to simply solve a 
 second order polynomial equation, and then tune $\psi_{\rm{M}}$ up to one of the
 two zeros of the quadratic function given by the r.-h.-s. of Eq.~(\ref{eq:DLMDt}). The two roots are  the 
 potentials of zero force   $\psi_{\rm{PZF}}.$
\par In the remainder of this sub-section, we analyze (\ref{eq:DLMDt}) in connection with two important limiting cases. 
\subsubsection{\label{sec:LSAM} Isolation Limit: Linear Superposition Approximation (LSA)}
\par Let us consider the situation in which the two microspheres are in near isolation:
$
\kappa L\gg1. \hspace{10pt}
$
In this limit, 
the double-layer around each microsphere is not affected by the presence of the neighboring one. Hence the potential is simply the sum of the solutions for isolated spheres. 
Accordingly, our exact results for the  linear system (\ref{eq:lsysXYt}) can be approximated to a much simpler form and solved analytically, as discussed in Appendix \ref{ap:C}.
The resulting force has a simple analytical form, and the coefficients
 appearing in  (\ref{eq:DLMDt}) are approximated by
\begin{eqnarray}
\mathcal{A}_{\rm D}\left(L\right)&\approx& \mathcal{C}_{\rm D}\left(L\right) \approx \mathcal{O}\left(e^{-2\kappa L}\right)\,\;\;\;\;\;\;\;(\kappa L\gg1)  \nonumber\\
\mathcal{B}_{\rm D}\left(L\right)&\approx&4\pi\epsilon_0\, \epsilon_m\frac{R_{1}}{\frac{R_{1}C_{I}}{\epsilon_0\, \epsilon_m}+1+\kappa R_{1}}\left(\frac{\bar{\sigma}_{1}}{\epsilon_0\, \epsilon_m}\right)  \nonumber\\
&&\times\left(1+ \kappa\mathcal{L}\right) \frac{R_{1}R_{2}}{\mathcal{L}^{2}}e^{-\kappa\left(\mathcal{L}-R_{1}-R_{2}\right)}\,,
\label{eq:ABCcoefDt}
\end{eqnarray}
which in turn simplifies (\ref{eq:DLMDt}) to 
\begin{eqnarray}
\rm{F}_{\rm {D}}(\mathit{L})&\approx&4\pi \epsilon_0\, \epsilon_m \psi_{\rm{M}}\psi_{01}\left(1+ \kappa\mathcal{L}\right) \frac{R_{1}R_{2}}{\mathcal{L}^{2}}e^{-\kappa\left(\mathcal{L}-R_{1}-R_{2}\right)}\,,\nonumber\\
&&
\label{eq:FDLSAt}
\end{eqnarray}
where $\psi_{01}$ is given by Eqs.~(\ref{psi01a}) and (\ref{psi01b}).
\par
Expression (\ref{eq:FDLSAt}) is the Bell and Levine double-layer force in the LSA limit \cite{bell70} and, as pointed out by \cite{trefalt16}, is independent of the dielectric microsphere boundary condition type (\ref{eq:CR}), since $\psi_{01}$ does not depend on the spherical regulation capacitance per unit area ${\rm C}^{\rm{sph}}_{I}$. From (\ref{eq:FDLSAt}), 
force suppresion in the LSA limit is obtained by simply grounding 
 the metallic sphere: 
\begin{equation}
 \psi_{\rm{PZF}}^{(\rm {LSA})}=0\,,
 \label{eq:PZFLSA}
\end{equation}
regardless of the dielectric electric potential value $\psi_{01}$.
\subsubsection{\label{sec:PFA}Proximity Force Approximation (PFA)}
\par Another  important  limit corresponds to the range of validity of the proximity force or Derjaguin approximation (PFA)  \cite{Derjaguin1934,blocki77,butt10,israelachvili11}, 
in which the distance $L$ and the Debye length $\lambda_{D}$ are both much smaller than the  radii:
\begin{eqnarray}
L&\ll&R_{1}\,, R_2 \label{eq:PFAcondition1}\,,\\
\lambda_{D}=\frac{1}{\kappa}&\ll& R_{1}\,, R_2\, .
\label{eq:PFAcondition2}
\end{eqnarray}
While (\ref{eq:PFAcondition1}) is a geometrical condition 
entailing that the sphere curvature is small at the scale of the distance between the surfaces,
 (\ref{eq:PFAcondition2}) in turn is a physical condition which assures that the interaction has a short range 
 when compared to the sphere diameter. Taken together, the two conditions allow one to 
 approximate the interaction between the microspheres by the
 average of the 
 interaction between two half-spaces (with planar interfaces) over the local distances.
  The resulting force is given by \cite{Derjaguin1934}
\begin{eqnarray}
\rm{F}_{\rm {D}}^{\rm (PFA)}(\mathit{L})&=&2\pi\left(\frac{R_{1}R_{2}}{R_{1}+R_{2}}\right)u_{\rm{planes}}(L)\,,
\label{eq:fPFA}
\end{eqnarray}
where $u_{\rm{planes}}(L)$ is the potential energy per unit of area between two half-spaces made of the same material as the corresponding spheres. As outlined in Appendix \ref{ap:D}, it is given by 
\begin{equation}
u_{\rm{planes}}(L)=-\int_{\infty}^{L}P\left(a\right)da\,,
\label{eq:uplanest}
\end{equation}
where $P\left(a\right)$ is the pressure on the dielectric half region surface. 

Eq.~(\ref{eq:fPFA}) 
leads to a quadratic function of $\psi_M$ as already discussed in connection with the exact solution and Eq.~(\ref{eq:DLMDt}). The corresponding coefficients
 $\mathcal{A}_{\rm D}^{\rm (PFA)}$,  $\mathcal{B}_{\rm D}^{\rm (PFA)}$, and  $\mathcal{C}_{\rm D}^{\rm (PFA)}$ are given  in Appendix \ref{ap:D}. 

In the particular case of 
the CC model, \emph{i.e.}
${\rm C}^{\rm{plane}}_{I}=0$,   it is possible to directly integrate (\ref{eq:uplanest}) and obtain
\begin{eqnarray}
\rm{F}_{\rm D}^{\rm (PFA)}(\mathit{L})&=&2\pi\left(\frac{R_{1}R_{2}}{R_{1}+R_{2}}\right)\times\nonumber\\
&&\left(\frac{\kappa\epsilon_0\, \epsilon_m e^{-\kappa L}\psi^{2}_{M}+2\sigma_{01}\psi_{\rm{M}}-\frac{\sigma^{2}_{01}}{\kappa\epsilon_0\, \epsilon_m}e^{-\kappa L}}{e^{\kappa L}+e^{-\kappa L}}\right)\,.\nonumber\\
&&\label{eq:FPFApsit}
\end{eqnarray}
This result leads to a simple analytical expression for the 
 PZF  in the proximity force approximation:
\begin{equation}
\psi_{\rm{PZF}}^{\rm (PFA)} = \frac{\sigma_{01}e^{\kappa L}}{\kappa\epsilon_0\, \epsilon_m} \left(-1 \pm \sqrt{1+ e^{-2\kappa L}} \right)\,.
\label{eq:PFAroots}
\end{equation} 
\subsection{\label{sec:examplesD}Examples}
\par
\begin{figure}
\centering
\includegraphics*[scale=0.4,angle=0]{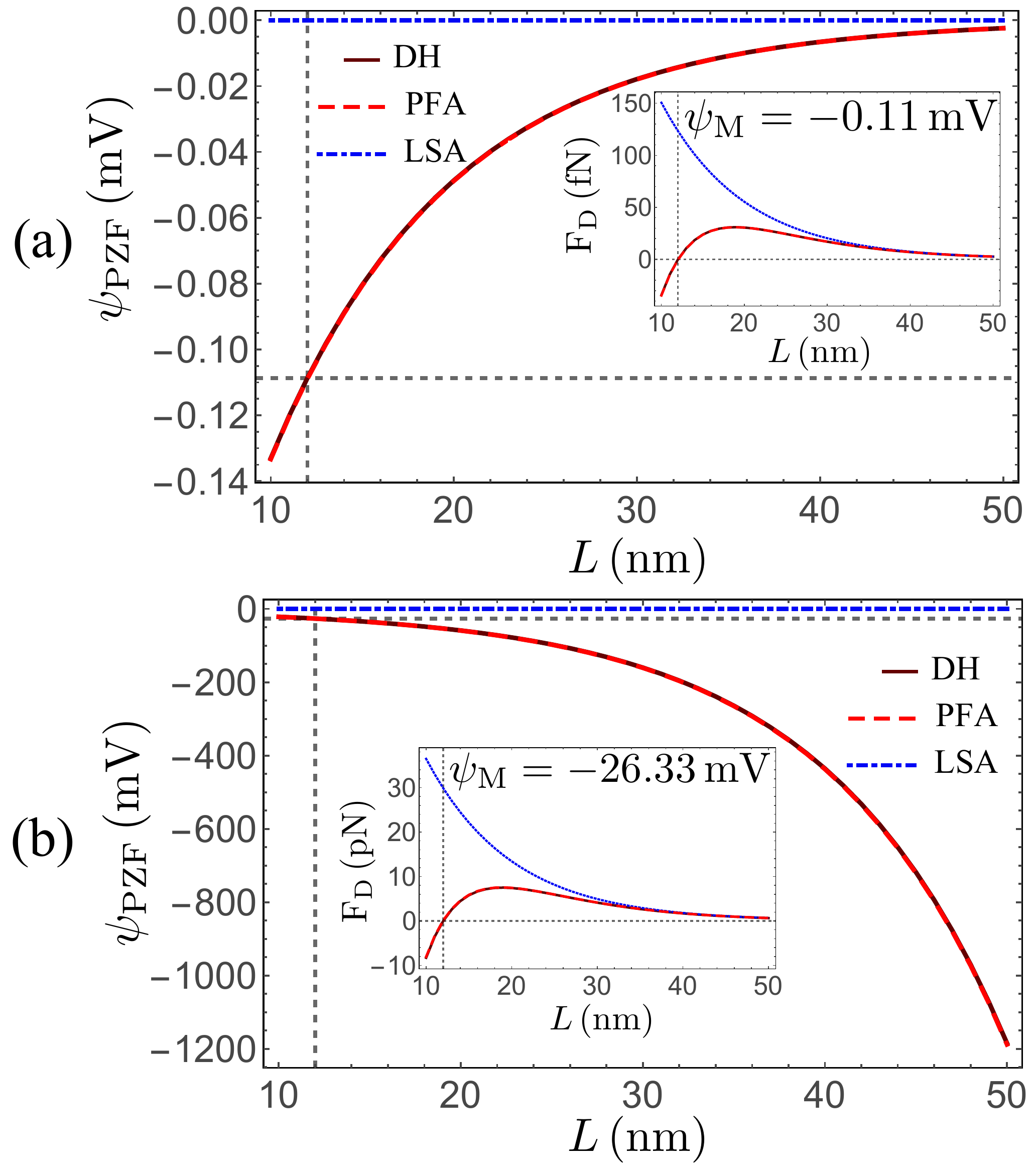}
\caption{(Color online) 
Variation of the potentials of zero force (PZF) with distance for the 
exact DH, LSA, and PFA models. 
We consider 
 a polystyrene microsphere (radius $R_{1}=10\,\mu \rm{m}$ and free charge density  $\sigma_{10}=-0.05\, \rm{mC/m}^{2}$) interacting with a metallic sphere of radius $R_{2}=15\,\mu \rm{m}.$ 
The Debye length is $\lambda_{\rm{D}}=10\,\rm{nm}.$
 DH and PFA results for $\psi_{\rm{PZF}}$ are consistent within $1\%$ for all the $L$ range considered in $(a)$, and within $2.5\%$ in $(b)$. Insets: double-layer force variation with distance
 for $\psi_{\rm{M}}=-0.11\,\rm{mV}$ in $(a)$ and $\psi_{\rm{M}}=-26.33\,\rm{mV}$ in $(b)$.}
\label{fig:3}
\end{figure}

 As our first example, let us consider a situation in which a polystyrene microsphere ($\epsilon_{p}=2.5$) of radius $R_{1}=10\,\mu \rm{m}$, free surface charge density $\sigma_{10}=-0.05\, \rm{mC/m}^{2}$ \cite{roberts07}, and $p=0.41$ (CR model) \cite{popa10} interacts with a metallic microsphere of $R_{2}=15\,\mu \rm{m}$ at an externally controlled $\psi_{\rm{M}}$. 
Both microspheres are immersed in water ($\epsilon_{m}=78.4$), where proper salt screening reduces the Debye length to $\lambda_{\rm{D}}=10\,\rm{nm}$. Fig.~\ref{fig:3} 
illustrates the  PZF roots 
for the exact DH calculation, as well as for the LSA and PFA.
We take $L$ varying from $10\,\rm{nm}$ to $50\,\rm{nm}.$

Each panel of Fig.~\ref{fig:3} shows one of the two roots of the r.-h.-s. of (\ref{eq:DLMDt}).
 For instance, if $L=12\,\rm{nm}$, the double-layer force is cancelled for either (a)  $\psi_{\rm{PZF}}=-0.11\,\rm{mV}$  or  (b) $\psi_{\rm{PZF}}=-26.33\,\rm{mV},$  
 as indicated by the horizontal and vertical dashed lines. 
 The corresponding inset plots show the force $\rm{F}_{\rm{D}}(\mathit{L})$ when $\psi_{\rm{M}}$ is tuned up to each of these values.
At such values, the force changes from repulsion (positive sign) to attraction (negative sign). The attraction results from the electric charge redistribution (electrostatic induction) on the metallic surface caused by the electric charges on the dielectric microsphere. The induced multipoles on the metallic sphere give rise to the well-known ``like-like attraction" behavior \cite{levin99}.

 In Fig.~\ref{fig:3}(a),  the PZFs  satisfy the DH linearity 
condition $\left|\psi_{\rm{PZF}}\right|\lesssim 25\,{\rm mV}$ for $T\sim300\,{\rm K}$ for the entire distance range shown in the plot. 
On the other hand, in  Fig.~\ref{fig:3}(b),
 the linear approximation is valid only in the region $L<20\,\rm{nm}.$ Thus, only one of the roots is consistent with the linear DH approximation leading to 
 (\ref{eq:DLMDt}) for  $L>20\,\rm{nm}.$
 Additionally, since we have $L/R_{1}<5\times10^{-3}$ and $\lambda_{\rm{D}}<L$ in this example, PFA conditions (\ref{eq:PFAcondition1}) and (\ref{eq:PFAcondition2}) are rather satisfied, and then the exact DH 
 and PFA predictions for the PZF are expected to agree. Indeed, we find a relative difference smaller than $1\%$ and $2.5\%$ in panels (a) and (b) of  Fig.~\ref{fig:3}, respectively. 
\par
\begin{figure}
\centering
\includegraphics*[scale=0.28]{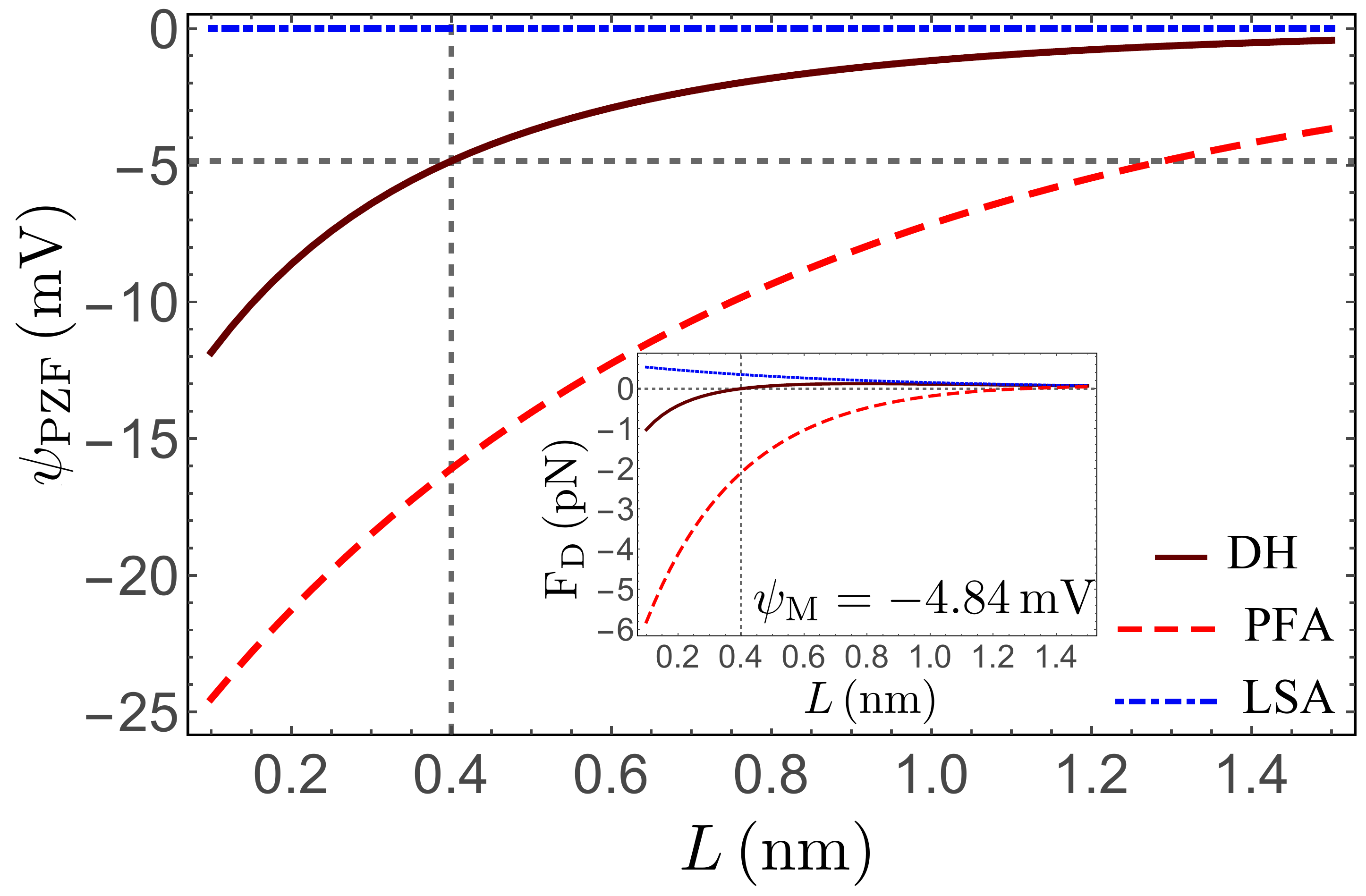}
\caption{(Color online) Same conventions as in Fig.~\ref{fig:3}, 
with a polystyrene microsphere of radius $R_{1}=0.5\,\mu \rm{m}$ and an electrolyte with  $\lambda_{\rm D}=750\,\rm{nm}.$
The exact DH and PFA predictions for $\psi_{\rm{PZF}}$  disagree even for distances smaller than $R_1.$ 
 Inset: double-layer force as a function of $L$ for $\psi_{\rm{M}}=-4.84\,\rm{mV}$. }
\label{fig:4}
\end{figure}

 As a second example, we consider a smaller dielectric microsphere and  a larger Debye length: $R_{1}=0.5\,\mu \rm{m}$ and $\lambda_{\rm D}=750\,\rm{nm},$ and keep all other parameters as in  Fig.~\ref{fig:3}.
In
Fig.~\ref{fig:4}, we plot the 
PZF root of smaller magnitude as a function of distance. 
Since the minimum distance $L$ varies from $0.1\,\mu \rm{m}$ up to $1.5\,\mu \rm{m}$,
 PFA conditions (\ref{eq:PFAcondition1}) and  (\ref{eq:PFAcondition2}) are not met. 
 For example, when $L=0.40\,\mu\rm{m}$, the exact DH and PFA models predict double-layer force suppression for very different PZF values, given by $\psi_{\rm{PZF}}=-4.84\,\rm{mV}$ and  
 $\psi_{\rm{PZF}}^{\rm (PFA)}=-16.10\,\rm{mV}$ respectively. 
 As a consequence, when setting $\psi_M=\psi_{\rm{PZF}}^{\rm (PFA)}$, the double-layer force is still significant,  $\rm{F}_{\rm{D}}(\mathit{L}=0.40\,\mu\rm{m})=0.76\,\rm{pN},$
 which could lead to a systematic error when measuring additional surface interactions like the Casimir force. 
Finally, we note that the LSA curves shown in the insets of Figs.~\ref{fig:3} and \ref{fig:4} fail to account for the crossover between repulsion to attraction as the distance decreases.
Moreover, LSA significantly overestimates the repulsive force for distances $L\stackrel{<}{\scriptscriptstyle\sim} 3\lambda_{\rm D}$ for the examples shown in Fig.~\ref{fig:3}.
\section{ \label{sec:isomet} Isolated metallic microsphere}

\subsection{\label{sec:model} Debye-H\"uckel theory with two metallic spherical surfaces}

\par 
In this section, we replace the dielectric sphere by a metallic one, which is  electrically isolated with total charge $Q.$
As in the previous section, 
we first solve the linear Poisson-Boltzmann equation (\ref{eq:lpbt}) giving the potential in the electrolyte medium. For that purpose, we take into account the appropriate boundary conditions.
 While the solutions outside the microspheres and inside the metallic microsphere 2 are still given by (\ref{eq:psiout12t}) and (\ref{eq:psiin2t}) respectively, the potential inside the isolated metallic microsphere (sphere 1)  is now given by 
\begin{eqnarray}
\psi^{I\!}(r_{1},\theta_{1})&=&\phi_{\rm{M}}\left(L\right)\,,
\label{eq:psiin1Mt}
\end{eqnarray} 
where $\phi_{\rm{M}}\left(L\right)$ varies with the distance $L$.
\par The boundary conditions at the surface of  the isolated metallic microsphere are given by
\begin{eqnarray}
\left.\psi^{I\!I\!I}(r_{1},\theta_{1})\right|_{r_1=R_1^{+}}&=&\phi_{\rm{M}}\left(L\right)\,,\label{eq:bcondmetaliso1}\\
-\left.\epsilon_{m}\vect{\hat{n}}\cdot\nabla\psi^{I\!I\!I}(r_{1},\theta_{1})\right|_{r_1=R_1^{+}}&=&\frac{\sigma_{1}\left(\theta_{1},\kappa L\right)}{\epsilon_{0}}\,,
\label{eq:bcondmetaliso2}
\end{eqnarray}
where the non-uniform distance-dependent 
surface charge density $\sigma_{1}$ is constrained by the condition that the total charge $Q$ is prescribed and distance independent. 
This type of boundary condition has been 
recently considered for the simpler planar geometry \cite{chan15,jackson99,hall75}.
 For the metallic microsphere surface held at $\psi_{\rm{M}}$, the boundary conditions are still given by (\ref{eq:bcondmetal1}) and (\ref{eq:bcondmetal2}).  
 
\par Following the method discussed in the previous section,  we derive the linear system of equations 
\begin{eqnarray}
&&\mathbf{X} + \amsmathbb{C}^{(1)} \cdot \mathbf{Y} = \phi_{\rm{M}} \, \mathbf{e}_0  \nonumber \\
&&\mathbf{Y} + \amsmathbb{C}^{(2)} \cdot \mathbf{X} = \psi_{\rm{M}} \, \mathbf{e}_0 
\label{eq:lsysXYIso}
\end{eqnarray}
where $e_{0i}= \delta_{i0}$,
\begin{eqnarray}
X_{j}\equiv  k_{j}(\kappa R_1) a_{j}& \hspace{10pt} , \hspace{10pt} Y_{j}\equiv k_{j}(\kappa R_2) b_{j}\,,
\label{eq:XYsysIso}
\end{eqnarray}
and
\begin{eqnarray}
\amsmathbb{C}^{(1)}_{jk}\equiv(2j+1)B_{jk}(\kappa \mathcal{L})\frac{i_{j}(\kappa R_1)}{k_k(\kappa R_2)} \nonumber \\
\amsmathbb{C}^{(2)}_{jk}\equiv(2j+1)B_{jk}(\kappa \mathcal{L})\frac{i_{j}(\kappa R_2)}{k_k(\kappa R_1)} \,.
\label{eq:CcondoM}
\end{eqnarray}
Solving this system, we then have
\begin{eqnarray}
&&X_{n} = \phi_{\rm{M}}(\amsmathbb{E}^{-1})_{n0} - \psi_{\rm{M}}\amsmathbb{F}^{(1)}_{n0}\,, \label{eq:solutionIsoX}\\
&&Y_{n} = \psi_{\rm{M}}\amsmathbb{G}_{n0}-\phi_{\rm{M}}{\amsmathbb{H}}_{n0} \,,
\label{eq:solutionIsoY}
\end{eqnarray}
where
\begin{eqnarray}
\amsmathbb{E} &\equiv& \mathbb{1} - \amsmathbb{C}^{(1)} \cdot \amsmathbb{C}^{(2)}\label{eq:EsysIso}\,,\\
\amsmathbb{G}&\equiv& \mathbb{1} + \amsmathbb{C}^{(2)} \cdot \amsmathbb{F}^{(1)}\label{eq:FsysIso}\,,\\
\amsmathbb{F}^{(1)} &\equiv& \amsmathbb{E}^{-1} \cdot \amsmathbb{C}^{(1)}\label{eq:GsysIso}\,,\\
\amsmathbb{H}&\equiv&  \amsmathbb{C}^{(2)} \cdot \amsmathbb{E}^{-1}\label{eq:HsysIso}\,.
\end{eqnarray}
\par Although (\ref{eq:solutionIsoX}) and (\ref{eq:solutionIsoY}) constitute the formal solution of the linear system (\ref{eq:lsysXYIso}), we actually do not know the potential $\phi_{\rm{M}}(L)$, which varies as a function of $L$. Indeed, since the total charge $Q$ of the isolated metallic microsphere is fixed, 
we should rewrite $\phi_{\rm{M}}(L)$  in terms of $Q$. By integrating 
$\sigma_{1}\left(\theta_{1},\kappa L\right)$ over the microsphere surface and 
using the orthogonality of the Legendre polynomials, 
we find 
\begin{equation}
X_{0}-\frac{\kappa R_{1}}{1+\kappa R_{1}}\sum_{k=0}^{\infty}\amsmathbb{C}^{\prime(1)}_{0k}Y_{k}=\frac{Q}{4\pi\epsilon_0\, \epsilon_m R_{1}\left(1+\kappa R_{1}\right)}\,,
\label{eq:phiMeq}
\end{equation}
where
\begin{equation}
\amsmathbb{C}^{\prime(1)}_{jk}\equiv (2j+1)B_{jk}(\kappa \mathcal{L})\frac{i^{\prime}_{j}(\kappa R_1)}{k_k(\kappa R_2)}\,.
\end{equation}
\par Combining this result with Eqs.~(\ref{eq:solutionIsoX}) and (\ref{eq:solutionIsoY}), we obtain
\begin{equation}
\phi_{\rm{M}}(L)=\Omega(L)\psi_{\rm{M}}+\Phi(L)\,,
\label{eq:phiMIso}
\end{equation}
where
\begin{eqnarray}
\Omega(L)&=&\frac{\amsmathbb{F}^{(1)}_{00} +\amsmathbb{K}_{00}}{\amsmathbb{E}^{-1}_{00}+\amsmathbb{L}_{00}}
\label{eq:omegaM}\,,\\
\Phi(L)&=&\frac{Q}{4\pi \epsilon_0\, \epsilon_m R_{1}\left(1+\kappa R_{1}\right)}\frac{1}{\amsmathbb{E}^{-1}_{00}+\amsmathbb{L}_{00}}
\label{eq:PhiM}\,,
\end{eqnarray}
and
\begin{eqnarray}
\amsmathbb{K}&=&\frac{\kappa R_{1}}{1+\kappa R_{1}}\amsmathbb{C}^{\prime(1)}\cdot\amsmathbb{G}\label{eq:Kiso}\,,\\
\amsmathbb{L}&=&\frac{\kappa R_{1}}{1+\kappa R_{1}} \amsmathbb{C}^{\prime(1)} \cdot \amsmathbb{H}\label{eq:Liso}\,.
\end{eqnarray}
Expression (\ref{eq:phiMIso}) relates the electrostatic potential $\phi_{\rm{M}}(L)$ with the known parameters $\psi_{\rm{M}}$ and $Q$.
\subsection{\label{sec:quadforceISO}DH double-layer force and Potential of Zero Force (PZF)}
\par Since a metallic microsphere surface at electrostatic equilibrium is an equipotential, the double-layer force calculation for this case is 
considerably simplified. Indeed, choosing $\partial\Re_{1}$ to be a spherical surface of radius  infinitesimally close to the surface of the isolated metallic sphere, and using (\ref{eq:thetatensorM}) combined with (\ref{eq:maxwellstresstensor}) and (\ref{eq:osmotic}),  we obtain for the double-layer force component along $\vect{\hat{z}}$
\begin{equation}
{\rm F}_{\rm M}=-\pi R^{2}_{1}\epsilon_0\, \epsilon_m\int_{-1}^{1}E^{2}_{r}\cos\theta_{1} \, d\!\cos\theta_{1}\,,
\label{eq:forcez1Iso}
\end{equation} 
 and 
\begin{eqnarray}
E_{r}=E^{I\!I\!I}_{r}(r_{1},\theta_{1}) = -\left.\frac{\partial\psi^{I\!I\!I}}{\partial r_{1}}(r_{1},\theta_{1})\right|_{r_{1}=R_{1}^{+}}\,.
\end{eqnarray}
Using (\ref{eq:psiout12t}), (\ref{eq:solutionIsoX}) and (\ref{eq:solutionIsoY}), together with (\ref{eq:phiMIso}), (\ref{eq:omegaM}) and (\ref{eq:PhiM}), we can then show that (\ref{eq:forcez1Iso}) can be rewritten as
\begin{equation}
\rm{F}_{\rm {M}}\left(\mathit{L}\right)=\mathcal{A}_{\rm {M}}\left(\mathit{L}\right)\psi^{2}_{M}+\mathcal{B}_{\rm {M}}\left(\mathit{L}\right)\psi_{\rm{M}}+\mathcal{C}_{\rm {M}}\left(\mathit{L}\right)\,,
\label{eq:DLMISOt}
\end{equation}
where the coefficients $\mathcal{A}_{\rm {M}}\left(\mathit{L}\right)$,  $\mathcal{B}_{\rm {M}}\left(\mathit{L}\right)$, and  $\mathcal{C}_{\rm {M}}\left(\mathit{L}\right)$ are given in Appendix \ref{ap:B}. Not surprisingly, as in (\ref{eq:DLMDt}), (\ref{eq:DLMISOt}) has again a simple quadratic dependence on $\psi_{\rm{M}}.$ 
As in the previous section, the PZF values are then given by the two roots of the r.-h.-s. of Eq.~(\ref{eq:DLMISOt}). 

\subsubsection{\label{sec:LSAPFAMISO}LSA and PFA Limits}
\par The LSA expression for the double-layer force in the current case can be obtained by using the same ideas discussed in section \ref{sec:LSAM}, 
leading to
\begin{eqnarray}
{\rm F}_{\rm M}(\mathit{L})\approx 4\pi\epsilon_0\, \epsilon_m \psi_{\rm{M}}\phi^{(0)}_{\rm{M}}\left(1+\kappa\mathcal{L}\right) \frac{R_{1}R_{2}}{\mathcal
{L}^{2}}e^{-\kappa\left(\mathcal{L}-R_{1}-R_{2}\right)}\,,\nonumber\\
&&
\label{eq:LSAMISO}
\end{eqnarray}
where
\begin{equation}
\phi^{(0)}_{\rm{M}}\equiv\frac{Q}{4\pi\epsilon_0\, \epsilon_m R_{1}(1+\kappa R_{1})}\,
\end{equation}
is the screened monopole contribution for the potential (\ref{eq:phiMIso}). Regardless of its value, it is clear that the LSA double-layer force vanishes for all distances if $\psi_{\rm{PZF}}^{(\rm {LSA})}=0$ as in (\ref{eq:FDLSAt}). Finally, let us also note that, in this isolation LSA limit, the surface charge density (\ref{eq:bcondmetal2}) of both metallic microspheres  are uniform and  given by
\begin{eqnarray}
\sigma_{01}&\approx&\frac{Q}{4\pi R^{2}_{1}}=\epsilon_0\, \epsilon_m\frac{1+\kappa R_{1}}{R_{1}}\phi^{(0)}_{\rm{M}}\,,\label{eq:sigma1iso}\\
\sigma_{02}&\approx&\epsilon_0\, \epsilon_m\frac{1+\kappa R_{2}}{R_{2}}\psi_{\rm{M}}\label{eq:sigma2iso}\,.
\end{eqnarray}
\par In order to calculate the force on the isolated metallic microsphere in the PFA approximation, we need the potential energy per unit area between two half regions: one held at an external electrostatic potential $\psi_{\rm{M}}$ and the other constrained to have a total charge $Q$. However, the latter condition combined with the planar symmetry (see Fig.\ref{fig:2}) lead to the stronger statement of a uniform (and distance independent) surface charge density $\sigma_{1}=Q/A$. Therefore, the force on the isolated metallic microsphere is given by (\ref{eq:FPFApsit}) with the simple substitution $\sigma_{01}\rightarrow\sigma_{1}$, leading to (see appendix \ref{ap:D})
\begin{eqnarray}
{\rm F}_{\rm M}^{\rm (PFA)}(\mathit{L})&=&2\pi\left(\frac{R_{1}R_{2}}{R_{1}+R_{2}}\right)\nonumber\\
&&\times\left(\frac{\kappa\epsilon_0\, \epsilon_m e^{-\kappa L}\psi^{2}_{M}+2\sigma_{1}\psi_{\rm{M}}-\frac{\sigma^{2}_{1}}{\kappa\epsilon_0\, \epsilon_m}e^{
-\kappa L}}{e^{\kappa L}+e^{-\kappa L}}\right)\,.\nonumber\\
&&\label{eq:FPFAMQt}
\end{eqnarray}
The PZF values are then simply given by
\begin{equation}
\psi_{\rm{PZF}}^{\rm (PFA)} = \frac{\sigma_1e^{\kappa L}}{\kappa\epsilon_0\, \epsilon_m} \left(-1 \pm \sqrt{1+ e^{-2\kappa L}} \right)\,.
\label{eq:PFArootsIso}
\end{equation} 
\subsection{\label{sec:examplesM}Examples}
\begin{figure*}
\centering
\includegraphics*[scale=0.6]{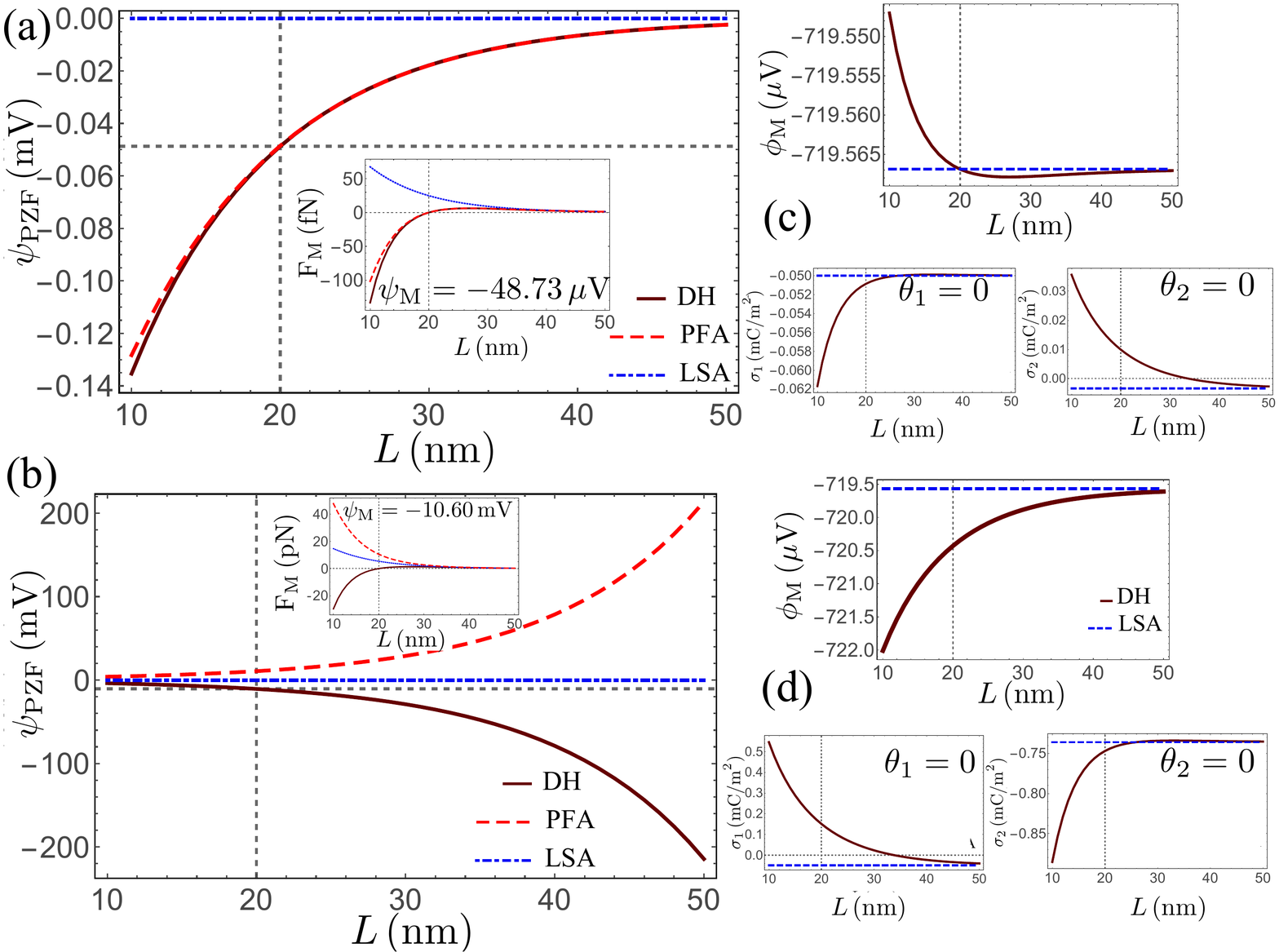}
\caption{(Color online) 
Same conventions as in Fig.~\ref{fig:3}, 
with an isolated metallic microsphere of radius $R_{1}=10\,\mu\rm{m}$
interacting with a metallic sphere of  radius $R_{2}=15\,\mu\rm{m}$ (at an externally controlled potential) across 
 an electrolyte with  $\lambda_{\rm D}=10\,\rm{nm}.$
Plots $(a)$ and $(b)$ show the two different  roots of Eq.~(\ref{eq:DLMISOt}). The PFA result is calculated for the value at isolation $\sigma_{1}=-0.05\, \rm{mC/m}^{2}.$
While the DH and the PFA predictions for the PZF are in good agreement in plot $(a)$, they are quite different in plot $(b).$ 
Insets show the  double-layer forces as function of distance for  $\psi_{\rm{M}}=-48.73\,\mu\rm{V}$ and $\psi_{\rm{M}}=-10.60\,\rm{mV}$. $(c)$ and $(d)$ illustrate the variation of the 
isolated microsphere (sphere 1)
electrostatic potential   with distance. They also show  the surface charge densities $\sigma_{1}$ and $\sigma_{2}$  evaluated at $\theta_{1}=0$ and $\theta_{2}=0,$ respectively. 
In $(c),$   $\sigma_{1}$ remains almost constant and equal to its PFA value, while $\sigma_{2}$ changes sign as a consequence of electrostatic induction occurring mainly in microsphere 2. 
This situation inverts for case $(d)$, where, in contrast, $\sigma_{1}$ changes significantly as electrostatic induction primarily happens in microsphere 1, whereas $\sigma_{2}$ remains almost equal to its isolation value. }
\label{fig:6}
\end{figure*}

 Let us consider the situation in which an isolated metallic microsphere of radius $R_{1}=10\,\mu\rm{m}$ and total charge $Q=-6.28\times10^{-14}\rm{C}$ interacts with another metallic microsphere of radius $R_{2}=15\,\mu\rm{m}$ at an externally controlled potential $\psi_{\rm{M}}$. 
They are both immersed in water, where proper salt screening reduces the Debye length to  $\lambda_{\rm{D}}=10\,\rm{nm}$. The two PZF roots of (\ref{eq:DLMISOt}) are shown separately in Figs.~\ref{fig:6}(a) and \ref{fig:6}(b) 
for $L$ varying from $10\,\rm{nm}$ to $50\,\rm{nm}$, together with the LSA and PFA predictions. 
The PFA is evaluated for the fixed surface charge density $\sigma_{1}=-0.05\, \rm{mC/m}^{2}$. For instance, the root shown in Fig.~\ref{fig:6}a values $\psi_{\rm{PZF}}=-48.73\,\mu\rm{V}$ at $L=20\,\rm{nm},$ as indicated by the horizontal and vertical dashed lines.
The inset shows that the double-layer force for  $\psi_M=-48.73\,\mu\rm{V}$ is indeed suppressed and changes sign at such distance. 

Fig.~\ref{fig:6}(a) shows an overall good agreement  between the DH and PFA values of the PZF, 
as expected  since PFA  conditions (\ref{eq:PFAcondition1}) and (\ref{eq:PFAcondition2}) are satisfied for the parameters corresponding to this figure.
On the other hand, for the second PZF root, shown in Fig.~\ref{fig:6}(b), DH and PFA predictions completely disagree, although the parameters are the same as in  Fig.~\ref{fig:6}(a). 
In order to understand such behavior, we plot 
 the variation of the isolated sphere electrostatic potential $\phi_{\rm{M}}$ with distance, together with the surface charge densities of both microspheres $\sigma_{1}$ and $\sigma_{2}$ evaluated at the points of closest separation $\theta_{1}=0$ and $\theta_{2}=0,$ in 
Fig. \ref{fig:6}(c) for the first root, and in Fig.~\ref{fig:6}(d) for the second one.

According to Fig. \ref{fig:6}(c), $\phi_{\rm{M}}$ is much larger than  the PZF values $\psi_{\rm{PZF}}$  displayed in Fig. \ref{fig:6}(a). 
As a consequence, electrostatic induction occurs mainly on the 
 sphere at the controlled potential $\psi_{\rm{PZF}}$ (sphere 2), making its surface charge density strongly distance dependent and non-uniform. 
 On the other hand, the isolated sphere surface charge density $\sigma_1$ is approximately uniform and distance independent, as also shown in Fig. \ref{fig:6}(c), thus explaining why the PFA 
 result is very close to the DH one. However,  some surface charge variation starts to build up on the isolated sphere  for  distances  below $20$ nm, making PFA slightly less accurate in this range. 

The roles of the two spheres are essentially interchanged when considering the PZF root shown in Fig.~\ref{fig:6}(b). 
In this case,  $\phi_{\rm{M}}$ is much smaller than $\psi_{\rm{PZF}}$ as shown in Fig. \ref{fig:6}(d), and electrostatic induction 
now occurs mainly on the isolated sphere, as indicated by the variations of $\sigma_1$ and $\sigma_2$ with distance. 
The strong variation of the isolated sphere charge density $\sigma_1$ is in striking contradiction with the assumptions associated to PFA, thus explaining why this approximation 
is unable to capture the qualitative features of the double-layer interaction in this case. For instance, 
 the DH model predicts $\psi_{\rm{PZF}}=-10.60\,\rm{mV}$ at $L=20\,\rm{nm}.$ The  double-layer force variation with distance for such potential  is illustrated in the inset of Fig.~\ref{fig:6}(b). The PFA force is always repulsive and in complete disagreement with the DH force, which displays a crossover from repulsion to attraction as the distance is decreased. Thus, we conclude that conditions  (\ref{eq:PFAcondition1}) and (\ref{eq:PFAcondition2}) are not sufficient for the validity of PFA when considering isolated metallic spheres. 
 
As a final remark, we note that the linearity condition is satisfied throughout most of the distance range ($L\stackrel{<}{\scriptscriptstyle\sim}40\,\rm{nm}$) shown in Fig.~\ref{fig:6}b. 
The nonlinear corrections for $40\,\rm{nm} \stackrel{<}{\scriptscriptstyle\sim} \mathit{L} < 50\,\rm{nm}$ should not affect the qualitative conclusions discussed here. 
%
%\begin{equation}
%L\,(\rm{nm})\,\,\,\,\,F_{\rm {D}}\,(\rm{fN})\,\,\,\,\,F_{\rm {D}}\,(\rm{pN})\,\,\,\,\,F_{\rm {M}}\,(\rm{fN})\,\,\,\,\,F_{\rm {M}}\,(\rm{pN})
%\end{equation}
%$\psi_{\rm{PZF}}\,(\rm{mV})$
%
%$\phi_{\rm{M}}\,(\mu\rm{V})$
%$\sigma_{1}\,(\rm{mC/m}^{2})$
%$\sigma_{2}\,(\rm{mC/m}^{2})$
%$\psi_{\rm{PZF}}=-48.73\,\mu\rm{V}$
%
\section{\label{sec:concpersp}Conclusion}
 We propose a protocol based on electrostatic potential  control to suppress the electrostatic double-layer forces between a metallic microsphere and a dielectric or another metallic microsphere.  The method works in a wide range of distances between the microspheres, beyond the PFA approximation, and even in cases where the PFA approximation is not well defined. Although the method relies on Poisson Boltzmann linearization, it is still accurate in several typical experimental conditions, as  illustrated  by a number of examples corresponding to realistic experimental parameters involving aqueous solutions. Additionally, the protocol is also suited to suppress the double-layer force in apolar media \cite{butt10,larson-smith12}, 
for which salt screening is usually not possible~\cite{sainis08,roberts08}. Finally, this method would be useful in force spectroscopy experiments probing additional interactions such as the Casimir force \cite{ether15}, particularly in salt unscreened scenarios.

\begin{acknowledgments}
The authors are grateful to C. Farina, R. de Melo e Souza,  M. Hartmann and G.-L. Ingold for helpful discussions. D.S.E., F.S.S.R., D.M.T., L.B.P., and P.A.M.N. would like to thank the Brazilian agencies
 National Council for Scientific and Technological Development (CNPq), the Coordination for the Improvement of Higher Education Personnel (CAPES), the National Institute of Science and Technology Complex Fluids (INCT-FCx), and the Research Foundations of S\~ao Paulo (FAPESP) and Rio de Janeiro (FAPERJ).
 D.S.E. and P.A.M.N. also acknowledge support from the PNPD CAPES-FAPERJ program. 
\end{acknowledgments}

\appendix
\section{\label{ap:A}The addition theorem for Bessel functions}
\par With the help of the addition theorem for Bessel functions \cite{langbein74,glendinning83,carnie93}, it is possible to rewrite expression (\ref{eq:lpbt}), the electric potential  $\psi^{I\!I\!I}(P)$ in the region outside the microspheres, in terms of variables related to a {\it single} coordinate system, that could be centered in either microsphere 1 (then the variables are $r_{1},\theta_{1}$) or microsphere 2 (and then we have $r_{2},\theta_{2}$):
\begin{eqnarray}
\psi^{I\!I\!I}(P)&=&\sum_{n=0}^{\infty}\left[a_{n}k_{n}(\kappa r_{1})P_{n}(\cos \theta_1)\right.\nonumber\\
&&\left.+b_{n}\sum_{m=0}^{\infty}(2m+1)B_{mn}(\kappa \mathcal{L})i_{m}(\kappa r_{1})P_{m}(\cos\theta_1)\right]\,,\nonumber\\
&=&\sum_{n=0}^{\infty}\left[a_{n}\sum_{m=0}^{\infty}(2m+1)B_{mn}(\kappa \mathcal{L})i_{m}(\kappa r_{2})P_{m}(\cos\theta_2)\right.\nonumber\\
&&+\left.b_{n}k_{n}(\kappa r_{2})P_{n}(\cos \theta_2)\right]\,,
\label{eq:psiout}
\end{eqnarray}
where 
\begin{equation}
B_{mn}(\kappa \mathcal{L})=\sum_{\nu=0}^{\infty}A_{mn}^{\nu}k_{m+n-2\nu}(\kappa \mathcal{L})\,,
\label{eq:Bcoef}
\end{equation} 
with
\begin{eqnarray}
A_{mn}^{\nu}&=&\frac{\Gamma(m-\nu+\frac{1}{2})\Gamma(n-\nu+\frac{1}{2})\Gamma(\nu+\frac{1}{2})}{\pi\Gamma(m+n-\nu+\frac{3}{2})}\times\nonumber\\
&&\frac{(m+n-\nu)!}{(m-\nu)!(n-\nu)!\nu!}\left(m+n-2\nu+\frac{1}{2}\right)\, .
\label{eq:Agamma}
\end{eqnarray} 
In these expressions, $\Gamma$ is the Gamma function, and $i_{n}$  and $k_{n}$ are the modified spherical Bessel function of the first and  third kind, respectively. 
They are defined as  
\begin{eqnarray}
 i_{n}(x)=(\pi/2x)^{1/2}I_{n+\frac{1}{2}}(x) \\
 k_{n}(x)=(\pi/2x)^{1/2}K_{n+\frac{1}{2}}(x)
 \label{eq:modsBessel}
\end{eqnarray}
where $I_n(x)$ and $K_n(x)$ are the modified cylindrical Bessel functions \cite{abramowitz65}.
\section{\label{ap:B}double-layer force coefficients}
In the expression (\ref{eq:DLMDt}) for the double-layer force between a metallic microsphere at electrostatic potential $\psi_{\rm{M}}$ and a charge regulated dielectric microsphere, the coefficients $\mathcal{A}_{\rm{D}}(L)$, $\mathcal{B}_{\rm{D}}(L)$ and $\mathcal{C}_{\rm{D}}(L)$ are given by
\begin{eqnarray}
\mathcal{A}_{\rm{D}}(\mathit{L})&\equiv&\pi\epsilon_0\, \epsilon_m\, \mathbf{\Lambda}\cdot\mathbb{\Gamma}\cdot\mathbf{\Lambda} \,,\nonumber\\
\mathcal{B}_{\rm{D}}(\mathit{L})&\equiv& \pi\epsilon_0\, \epsilon_m \, \left(\mathbf{\Lambda} \cdot \mathbb{\Gamma} \cdot \mathbf{\Xi} +  \mathbf{\Xi}\cdot \mathbb{\Gamma} \cdot\mathbf{\Lambda}\right) \nonumber \\
&&\hspace{-20pt}+\frac{4\pi}{3}R_{1}\bar{\sigma}_{1}\left(\frac{R_{1}\rm{C}^{\rm{sph}}_{I}}{\epsilon_0\, \epsilon_m } +\frac { \epsilon_{p}}{\epsilon_{m}}+2\right) \mathbf{e}^{\rm{T}}_1 \cdot \mathbf{\Lambda} \,, \nonumber \\
\mathcal{C}_{\rm{D}}(\mathit{L}) &\equiv& \pi\epsilon_0\, \epsilon_m\,\mathbf{\Xi}\cdot\mathbb{\Gamma}\cdot\mathbf{\Xi}  \nonumber \\&& \hspace{-20pt} + \frac{4\pi}{3}R_{1}\bar{\sigma}_{1}\left(\frac{R_{1}\rm{C}^{\rm{sph}}_{I}}{\epsilon_0\, \epsilon_m}
+\frac{\epsilon_{p}}{\epsilon_{m}}+2\right)\mathbf{e}^{\rm{T}}_1\cdot\mathbf{\Xi}\,,\nonumber \\
&&\label{eq:ABCcoefD}
\end{eqnarray}
where $e^{\rm{T}}_{1i} = \delta_{1i}$,
\begin{eqnarray}
\Lambda_{k} &\equiv& -\frac{k_k(\kappa R_1)}{A_k} \amsmathbb{F}_{k0} +\sum_{l=0}^{\infty}\amsmathbb{B}^{\prime}_{kl}\amsmathbb{G}_{l0}\,, \label{eq:Lambda}\\
\Xi_{k} &\equiv&  \frac{R_1 \bar{\sigma}_1}{\epsilon_0\, \epsilon_m} \left[\frac{k_k(\kappa R_1)}{A_k} \left(\amsmathbb{E}^{-1}\right)_{k0} - \sum_{l=0}^{\infty}\amsmathbb{B}^{\prime}_{kl}\amsmathbb{H}_{l0}\right]\,,\nonumber\\
&&\label{eq:Xi}
\end{eqnarray}
being
\begin{eqnarray}
\amsmathbb{B}^{\prime}_{kl}&\equiv&(2k+1)\frac{B_{kl}(\kappa \mathcal{L})}{k_l(\kappa R_2)}i_{k}(\kappa R_{1})\,.
\label{eq:Bbarprime}
\end{eqnarray}
Furthermore, $\mathbb{\Gamma}$ is a matrix which its elements are independent of the distance $L$ and are given by 
\begin{eqnarray}
\mathbb{\Gamma}_{ij}&\equiv&\left\{\left(\kappa R_{1}\right)^{2}\right.\nonumber\\
&&\left.-\left[i\left(\frac{\epsilon_{p}}{\epsilon_{m}}\right)+\frac{R_{1}{\rm C}^{\rm{sph}}_{I}}{\epsilon_0\, \epsilon_m}\right]\left[j\left(\frac{\epsilon_{p}}{\epsilon_{m}}\right)+\frac{R_{1}{\rm C}^{\rm{sph}}_{I}}{\epsilon_0\, \epsilon_m}\right]\right\}\amsmathbb{P}^{(1)}_{ij}\nonumber\\
&&-2\left[i\left(\frac{\epsilon_{p}}{\epsilon_{m}}\right)+\frac{R_{1}{\rm C}^{\rm{sph}}_{I}}{\epsilon_0\, \epsilon_m}\right]\amsmathbb{P}^{(2)}_{ij}+\amsmathbb{P}^{(3)}_{ij}\,,
\label{eq:Gammacoeff}
\end{eqnarray}
where
\begin{eqnarray}
\amsmathbb{P}^{(1)}_{ij}&=&\int_{-1}^{1}\mu P_{i}(\mu)P_{j}(\mu)d\mu\nonumber\\
&=&\left\{\begin{array}{ll}\frac{2(j+1)}{(2j+3)(2j+1)}, &\mbox{$i=j+1$}\\
\frac{2j}{(2j+1)(2j-1)},&\mbox{$i=j-1$}\\
0,&\mbox{$i\neq j\pm 1$}
\label{eq:integralP1}
\end{array}
\right.
\end{eqnarray}
\begin{eqnarray}
\amsmathbb{P}^{(2)}_{ij}&=&\int_{-1}^{1}\left(1-\mu^{2}\right)P_{i}(\mu)P^{\prime}_{j
}(\mu)d\mu\nonumber\\
&=&\left\{\begin{array}{ll}-\frac{2j(j+1)}{(2j+3)(2j+1)}, &\mbox{$i=j+1$}\\
\frac{2j(j+1)}{(2j+1)(2j-1)},&\mbox{$i=j-1$}\\
0,&\mbox{$i\neq j\pm 1$}
\end{array}
\right.
\label{eq:integralP2}
\end{eqnarray}
\begin{eqnarray}
\amsmathbb{P}^{(3)}_{ij}&=&\int_{-1}^{1}\mu\left(1-\mu^{2}\right)P^{\prime}_{i}
(\mu)P^{\prime}_{j}(\mu)d\mu \nonumber\\
&=&\left\{\begin{array}{ll}\frac{2j(j+1)(j+2)}{(2j+3)(2j+1)}, &\mbox{$i=j+1$}\\
\frac{2j(j-1)(j+1)}{(2j+1)(2j-1)},&\mbox{$i=j-1$}\\
0,&\mbox{$i\neq j\pm 1$}
\label{eq:integralP3}
\end{array}
\right.
\end{eqnarray}
\par For the case of a metallic microsphere at electrostatic potential $\psi_{\rm{M}}$ and an isolated metallic microsphere with total charge $Q$, the coefficients $\mathcal{A}_{\rm{M}}(\mathit{L})$, $\mathcal{B}_{\rm{M}}(\mathit{L})$ and $\mathcal{C}_{\rm{M}}(\mathit{L})$ in (\ref{eq:DLMISOt}) are given by
\begin{eqnarray}
\mathcal{A}_{\rm {M}}\left(\mathit{L}\right)&\equiv&\mathcal{E}_{\rm {M}}(\mathit{L})+\Omega\left(L\right)\mathcal{F}_{\rm {M}}(\mathit{L})+\Omega^{2}\left(L\right)\mathcal{G}_{\rm {M}}(\mathit{L})\,,\nonumber \\
\mathcal{B}_{\rm {M}}\left(\mathit{L}\right)&\equiv&\Phi\left(\mathit{L}\right)\mathcal{F}_{\rm {M}}(\mathit{L})+2\,\Omega\left(L\right)\Phi\left(\mathit{L}\right)\mathcal{G}_{\rm {M}}(\mathit{L})\,,\nonumber \\
\mathcal{C}_{\rm {M}}\left(\mathit{L}\right)&\equiv&\Phi^{2}\left(\mathit{L}\right)\mathcal{G}_{\rm {M}}(\mathit{L})\,,
\label{eq:EFGM}
\end{eqnarray}
where
\begin{widetext}
\begin{eqnarray}
\mathcal{E}_{\rm {M}}(\mathit{L})&\equiv& -\pi\epsilon_0\, \epsilon_m(\kappa R_{1})^2\mathbf{e}^{\rm{T}}_{0}\cdot\left[\tilde{\amsmathbb{F}}^{(1)\rm{T}}\cdot \amsmathbb{M}^{(1)}\cdot \tilde{\amsmathbb{F}}^{(1)}-2\,\tilde{\amsmathbb{G}}^{\rm{T}}\cdot\tilde{\amsmathbb{B}}\cdot \amsmathbb{M}^{(2)}\cdot\tilde{\amsmathbb{F}}^{(1)}+\tilde{\amsmathbb{G}}^{\rm{T}}\cdot\tilde{\amsmathbb{B}}\cdot \amsmathbb{M}^{(3)}\cdot\tilde{\amsmathbb{B}}\cdot\tilde{\amsmathbb{G}}\right]\cdot\mathbf{e}_0\,,\\
\mathcal{F}_{\rm {M}}(\mathit{L})&\equiv& -\pi\epsilon_0\, \epsilon_m(\kappa R_{1})^2\mathbf{e}^{\rm{T}}_{0}\cdot\left[-\tilde{\amsmathbb{E}}^{-1\rm{T}}\cdot \amsmathbb{M}^{(1)}\cdot \tilde{\amsmathbb{F}}^{(1)}-\tilde{\amsmathbb{F}}^{(1)\rm{T}}\cdot \amsmathbb{M}^{(1)}\cdot \tilde{\amsmathbb{E}}^{-1}+2\,\tilde{\amsmathbb{G}}^{\rm{T}}\cdot\tilde{\amsmathbb{B}}\cdot \amsmathbb{M}^{(2)}\cdot\tilde{\amsmathbb{E}}^{-1}+2\,\tilde{\amsmathbb{H}}^{\rm{T}}\cdot\tilde{\amsmathbb{B}}\cdot \amsmathbb{M}^{(2)}\cdot\tilde{\amsmathbb{F}}^{(1)}\right.\nonumber\\
&&\left.-\tilde{\amsmathbb{G}}^{\rm{T}}\cdot\tilde{\amsmathbb{B}}\cdot \amsmathbb{M}^{(3)}\tilde{\amsmathbb{B}}\cdot\tilde{\amsmathbb{H}}-\tilde{\amsmathbb{H}}^{\rm{T}}\cdot\tilde{\amsmathbb{B}}\cdot \amsmathbb{M}^{(3)}\cdot\tilde{\amsmathbb{B}}\cdot\tilde{\amsmathbb{G}}\right]\cdot\mathbf{e}_0\,,\\
\mathcal{F}_{\rm {M}}(\mathit{L})&\equiv& -\pi\epsilon_0\, \epsilon_m(\kappa R_{1})^2\mathbf{e}^{\rm{T}}_{0}\cdot\left[\tilde{\amsmathbb{E}}^{-1\rm{T}}\cdot \amsmathbb{M}^{(1)}\cdot \tilde{\amsmathbb{E}}^{-1}-2\,\tilde{\amsmathbb{H}}^{\rm{T}}\cdot\tilde{\amsmathbb{B}}\cdot \amsmathbb{M}^{(2)}\cdot\tilde{\amsmathbb{E}}^{-1}+\tilde{\amsmathbb{H}}^{\rm{T}}\cdot\tilde{\amsmathbb{B}}\cdot \amsmathbb{M}^{(3)}\cdot\tilde{\amsmathbb{B}}\cdot\tilde{\amsmathbb{H}}\right]\cdot\mathbf{e}_0\,,
\end{eqnarray}
\end{widetext} 
where
\begin{eqnarray}
\amsmathbb{M}^{(1)}_{ij}&\equiv& k^{\prime}_{i}(\kappa R_{1}) k^{\prime}_{j}(\kappa R_{1})\amsmathbb{P}^{(1)}_{ij}\,, \\
\amsmathbb{M}^{(2)}_{ij}&\equiv& i^{\prime}_{i}(\kappa R_{1}) k^{\prime}_{j}(\kappa R_{1})\amsmathbb{P}^{(1)}_{ij}\,, \\
\amsmathbb{M}^{(3)}_{ij}&\equiv& i^{\prime}_{i}(\kappa R_{1}) i^{\prime}_{j}(\kappa R_{1})\amsmathbb{P}^{(1)}_{ij}\,, 
\end{eqnarray}
and
\begin{eqnarray}
\tilde{\amsmathbb{B}}_{ij}&\equiv& (2j+1) B_{ij}\,,\\
\tilde{\amsmathbb{E}}_{ij}&\equiv& \frac{\amsmathbb{E}_{ij}}{k_{i}\left(\kappa R_{1}\right)}\,,\\
\tilde{\amsmathbb{G}}_{ij}&\equiv& \frac{\amsmathbb{G}_{ij}}{k_{i}\left(\kappa R_{2}\right)}\,,\\
\tilde{\amsmathbb{F}}^{(1)}_{ij}&\equiv& \frac{\amsmathbb{F}^{(1)}_{ij}}{k_{i}\left(\kappa R_{1}\right)}\,,\\
\tilde{\amsmathbb{H}}_{ij}&\equiv& \frac{\amsmathbb{H}_{ij}}{k_{i}\left(\kappa R_{2}\right)}\,.
\end{eqnarray}
\section{\label{ap:C} Linear Superposition Aproximation (LSA)}
\par To solve the linear system of equations (\ref{eq:lsysXYt}) in the LSA limit, let us firstly note that (\ref{eq:Bcoef}) can be written as 
\begin{eqnarray}
B_{nm}(\kappa 
\mathcal{L})&=&\frac{\pi}{2\kappa\mathcal{L}}e^{-\kappa\mathcal{L}}\sum_{\nu=0}^
{\infty}A_{nm}^{\nu}R(n+m-2\nu+\frac{1}{2},\kappa \mathcal{L})\,,\nonumber\\
&&
\label{eq:Bexpgen}
\end{eqnarray}
where we have used the expansion for the third order modified spherical Bessel function \cite{abramowitz65}. Retaining only the $k=0,1$ terms, it is possible to show that 
\begin{eqnarray}
B_{00}(\kappa\mathcal{L})&\approx&\frac{\pi}{2\kappa\mathcal{L}}e^{
-\kappa\mathcal{L}}\,,\label{eq:B00}\\
B_{01}(\kappa \mathcal{L})&\approx&\frac{\pi(1+\kappa \mathcal{L})}{2(\kappa 
\mathcal{L})^{2}}e^{-\kappa \mathcal{L}}\label{eq:B01}\,.
\end{eqnarray}
(\ref{eq:Bexpgen}), together with (\ref{eq:Bcoeff}) and (\ref{eq:Ccoeff}), shows that the second term in (\ref{eq:Esys}) has order $\mathcal{O}\left(e^{-2\kappa  \mathcal{L}}\right)$, and therefore can be neglected when compare to terms of order $\mathcal{O}\left(e^{-\kappa  \mathcal{L}}\right)$. As a result, (\ref{eq:Esys}) reduces to
\begin{eqnarray}
\amsmathbb{E}_{ij}\approx\amsmathbb{E}^{-1}_{ij}\approx\delta_{ij}\,,
\label{eq:Eapprox}
\end{eqnarray}
allowing us to write (\ref{eq:Fsys}), (\ref{eq:Gsys}) and (\ref{eq:Hsys}) as
\begin{eqnarray}
\amsmathbb{F}_{ij}&\approx&\amsmathbb{B}_{ij}\,,\nonumber\\
\amsmathbb{G}_{ij}&\approx&\delta_{ij}\,,\nonumber\\
\amsmathbb{H}_{ij}&\approx&\amsmathbb{C}_{ij}\,.
\label{eq:FGHapprox}
\end{eqnarray}
Using these results in (\ref{eq:Lambda}) and (\ref{eq:Xi}), we then find
\begin{eqnarray}
\mathbb{\Lambda}_{k} \left(L\right) &\equiv& -\frac{k_k(\kappa R_1)}{A_k} \amsmathbb{B}_{k0} +\amsmathbb{B}^{\prime}_{k0}\,, \label{eq:LambdaLSA}\\
\Xi_{k} \left(L\right) &\equiv& \frac{R_1 \bar{\sigma}_1}{\epsilon_0\, \epsilon_m} \frac{k_k(\kappa R_1)}{A_k} \delta_{k0}\,.\label{eq:XiLSA}
\end{eqnarray}
Substituting this expressions into (\ref{eq:ABCcoefD}), using (\ref{eq:Bbarprime}) and (\ref{eq:Gammacoeff}), and making lengthy but straightforward calculations, we are lead to (\ref{eq:ABCcoefDt}).
\section{\label{ap:D} Proximity Force Approximation (PFA)}
\par
\begin{figure}[h]
\centering
\includegraphics*[scale=0.28]{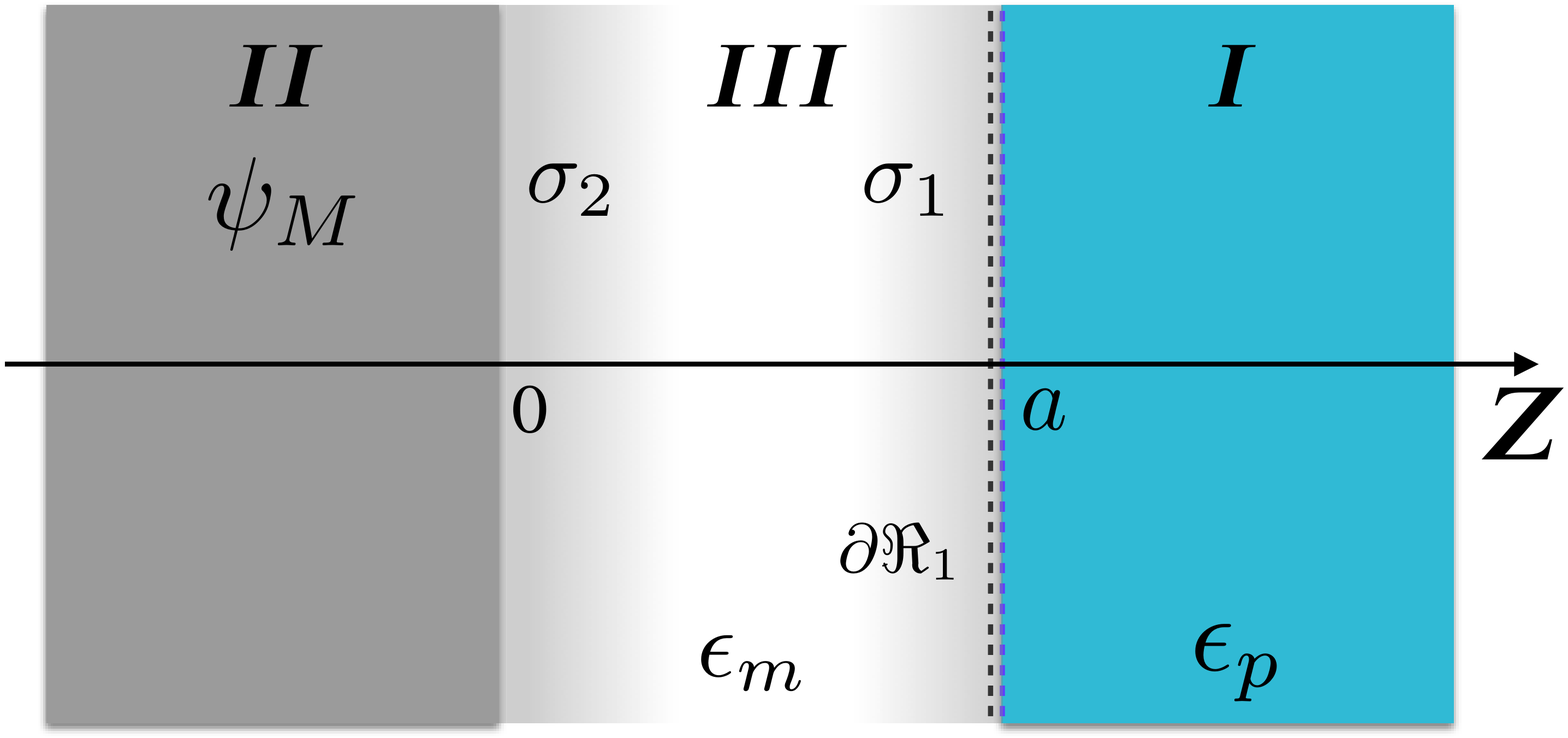}
\caption{Dielectric half region 1 and metallic half region 2 at $\psi_{\rm{M}}$.}
\label{fig:2}
\end{figure}

 Consider two half regions region $I$ and $I\!I$, as illustrated in Fig. \ref{fig:2}, being the first dielectric with relative permittivity $\epsilon_{p}$ and the second metallic at an externally controlled potential $\psi_{\rm{M}}$. The dielectric plane may exchange charge with the medium by ion adsorption or dissociation processes \cite{butt10,russel89,trefalt16}, being $\sigma_{1}$ its surface free charge density. Again, $Z:Z$ electrolytes are dissolved in region $I\!I\!I$ between them and are in thermal equilibrium with the bath. The half regions are separated by the distance $a$.
\par To calculate the interaction potential energy per unit area $u_{\rm{planes}}(L)$ for this configuration, we must solve the boundary value problem defined by the equations
\begin{eqnarray}
\frac{d^{2}\psi^{I\!I\!I}}{dz^{2}}(z)&=&\kappa^{2}\psi^{I\!I\!I}(z),\,\,\,0\leq 
z \leq a\,,\nonumber\\
 \psi^{I\!I}(z)&=&\psi_M,\,\,\,z<0 \,,\nonumber\\
\frac{d^{2}\psi^{I}}{dz^{2}}(z)&=&0,\,\,\,z>a\,.
\label{eq:PFAequations}
\end{eqnarray}
subject to the boundary conditions at $z=0$ and $z=a$
\begin{eqnarray}
\left.\psi^{I\!I}(z)\right|_{z=0^{-}}&=&\left.\psi^{I\!I\!I}(z)\right|_{z=0^{+}}\hspace{10pt}\,,\label{eq:bcPFA1a}\\
-\epsilon_{m}\left.\frac{d\psi^{I\!I\!I}}{dz}(z)\right|_{z=0^{+}}&=&\frac{\sigma_{2}(\kappa a)}{\epsilon_{0}}\,,\label{eq:bcPFA1b}\\
\left.\psi^{I\!I\!I}(z)\right|_{z=a^{-}}&=&\left.\psi^{I}(z)\right|_{z=a^{+}}\hspace{10pt}\,,\label{eq:bcPFA2a}\\
\epsilon_{m}\left.\frac{d\psi^{I\!I\!I}}{dz}(z)\right|_{z=a^{-}}-\epsilon_{p}\left.\frac{d\psi^{I}}{dz}(z)\right|_{z=a^{+}}&=&\frac{\sigma_{1}(\kappa 
a)}{\epsilon_{0}}\,,\label{eq:bcPFA2b}
\end{eqnarray}
where 
\begin{eqnarray}
\sigma_{1}(\kappa a)&=&\sigma_{01}-{\rm C}^{\rm{plane}}_{I}\left[\psi^{I\!I\!I}(\kappa a)-\mbox{C}\left(\infty\right)\right]\,,\nonumber\\
&=&\bar{\sigma}_{1}-{\rm C}^{\rm{plane}}_{I}\psi^{I\!I\!I}(\kappa a)\,,
\label{eq:sigmabarPFA}
\end{eqnarray}
being
\begin{equation}
\bar{\sigma}_{1}=\sigma_{01}+{\rm C}^{\rm{plane}}_{I} \mbox{C}\left(\infty\right)\,.
\label{eq:sigmabarPFA}
\end{equation}
Similarly to (\ref{eq:CR}), $\sigma_{01}$ and $ \mbox{C}\left(\infty\right)$ are, respectively, the surface charge density and electrostatic potential of the dielectric half region $I$ in near isolation, and the constant ${\rm C}^{\rm{plane}}_{I}=-\partial\sigma_{1}/\partial\psi\geq0$, named plane regulation capacitance per unit area (or plane inner layer capacitance per unit area), quantifies the dielectric plane surface dissociation rates and it is assumed to be distance independent. Again, we may define a parameter $p$, such that
\begin{equation}
p=\frac{{\rm C}^{\rm{plane}}_{D}}{{\rm C}^{\rm{plane}}_{D}+{\rm C}^{\rm{plane}}_{I}}\,,
\label{eq:plane}
\end{equation}
where
\begin{equation}
{\rm C}^{\rm{plane}}_{D}=\kappa\epsilon_0\, \epsilon_m
\end{equation}
is the diffuse layer capacitance per unit area of the isolated plane 1. Note that for a correct comparison between a CR dielectric microsphere and a CR dielectric half region, (\ref{eq:sigmabarPFA}) and (\ref{eq:plane}) above must be, respectively, equal to (\ref{eq:CR}) and (\ref{eq:psphere}).
\par The general solutions for the potential in the three regions are respectively given by
\begin{eqnarray}
\psi^{I\!I\!I}(z)&=&\mbox{A}e^{\kappa z}+\mbox{B}e^{-\kappa z},\,\,\,0\leq z \leq a\,,\nonumber\\
\psi^{I\!I}(z)&=&\psi_{\rm{M}},\hspace{60pt} z<0\,,\nonumber\\
\psi^{I}(z)&=&\mbox{C},\hspace{68pt} z>a\,.
\label{eq:planesol}
\end{eqnarray}
Using the boundary conditions (\ref{eq:bcPFA1a}) and (\ref{eq:bcPFA2b}), we then get
\begin{eqnarray}
\mbox{A}&=&\frac{\frac{\bar{\sigma}_{1}}{\kappa\epsilon_0\, \epsilon_m}+\left(1-\frac{{\rm C}^{\rm{plane}}_{I}}{\kappa\epsilon_0\, \epsilon_m}\right)\beta\psi_{\rm{M}}}{\left(1+\frac{{\rm C}^{\rm{plane}}_{I}}{\kappa\epsilon_0\, \epsilon_m}\right)\alpha+\left(1-\frac{{\rm C}^{\rm{plane}}_{I}}{\kappa\epsilon_0\, \epsilon_m}\right)\beta}\,,\label{eq:planeA}\\
\mbox{B}&=&\frac{-\frac{\bar{\sigma}_{1}}{\kappa\epsilon_0\, \epsilon_m}+\left(1+\frac{{\rm C}^{\rm{plane}}_{I}}{\kappa\epsilon_0\, \epsilon_m}\right)\alpha\psi_{\rm{M}}}{\left(1+\frac{{\rm C}^{\rm{plane}}_{I}}{\kappa\epsilon_0\, \epsilon_m}\right)\alpha+\left(1-\frac{{\rm C}^{\rm{plane}}_{I}}{\kappa\epsilon_0\, \epsilon_m}\right)\beta}\label{eq:planeB}\,,
\end{eqnarray}
where $\alpha\equiv e^{\kappa a}$, $\beta\equiv e^{-\kappa a}.$ For consistence, substituting (\ref{eq:planeA}) and (\ref{eq:planeB}) in (\ref{eq:bcPFA2a}), we find
\begin{eqnarray}
\mbox{C}\left(a\right)&=&\alpha\mbox{A}+\beta\mbox{B}\nonumber\\
&=&\frac{\left(\alpha-\beta\right)\frac{\bar{\sigma}_{1}}{\kappa\epsilon_0\, \epsilon_m}+2\psi_{\rm{M}}}{\left(1+\frac{{\rm C}^{\rm{plane}}_{I}}{\kappa\epsilon_0\, \epsilon_m}\right)\alpha+\left(1-\frac{{\rm C}^{\rm{plane}}_{I}}{\kappa\epsilon_0\, \epsilon_m}\right)\beta}\,.
\end{eqnarray}
 Taking the limit $a\rightarrow\infty$, we then find 
 \begin{equation}
 \mbox{C}\left(\infty\right)\rightarrow\frac{\bar{\sigma}_{1}}{\kappa\epsilon_0\, \epsilon_m\left(1+\frac{{\rm C}^{\rm{plane}}_{I}}{\kappa\epsilon_0\, \epsilon_m}\right)}.
 \end{equation}
Using (\ref{eq:sigmabarPFA}) in the above expression and solving for $\mbox{C}\left(\infty\right)$, we finally get
\begin{equation}
 \mbox{C}\left(\infty\right)=\frac{\sigma_{01}}{\kappa\epsilon_0\, \epsilon_m}\,,
\end{equation}
which is the electrostatic potential for the dielectric half region in isolation\cite{butt10}.
\par The above results enable us to calculate the pressure on the surface $\partial\mathcal{R}_{1}$ due to the adsorbed electrical charges in plane $z=0$ and the medium electrolytes. In fact, according to (\ref{eq:F1}), 
\begin{eqnarray}
\left.P\right|_{z=a^{-}}&=&\vect{\hat{z}}\cdot\overleftrightarrow{\vect{\Theta}}\cdot\left(-\vect{\hat{z}}\right)\,,\nonumber\\
&=&\frac{\epsilon_0\, \epsilon_m\kappa^{2}}{2}\psi^{2}-\frac{\epsilon_0\, \epsilon_m}{2}\left(\frac{d\psi}{dz}\right)^{2}\,,
\label{eq:pressurePFA}
\end{eqnarray} 
where we have omitted the index $I\!I\!I\!$ for simplicity and we used the fact that electrostatic potential $\psi$ depends only upon $\mbox{z}$. Using (\ref{eq:planeA}) and (\ref{eq:planeB}) in (\ref{eq:planesol}) and the result in (\ref{eq:pressurePFA}), we then have 
\begin{widetext}
\begin{eqnarray}
P(a) &=& 2\epsilon_0\, \epsilon_m\kappa^{2}\mbox{A}\mbox{B} \nonumber\\
&=& \frac{2}{\epsilon_0\, \epsilon_m}
\left\{\frac{\left(\kappa\epsilon_0\, \epsilon_m\right)^{2}\left[
1-\left(\frac{{\rm C}^{\rm{plane}}_{I}}{\kappa\epsilon_0\, \epsilon_m}\right)^{2}\right]\psi^{2}_{\rm{M}}+\kappa\epsilon_0\, \epsilon_m\bar{\sigma}_{1}\left[\left(1+\frac{{\rm C}^{\rm{plane}}_{I}}{\kappa\epsilon_0\, \epsilon_m}\right)e^{\kappa a}-\left(1-\frac{{\rm C}^{\rm{plane}}_{I}}{\kappa\epsilon_0\, \epsilon_m}\right)e^{-\kappa a}\right]\psi_{\rm{M}}-\bar{\sigma}_{1}^{2}}{\left[\left(1+\frac{{\rm C}^{\rm{plane}}_{I}}{\kappa\epsilon_0\, \epsilon_m}\right)e^{\kappa a}+\left(1-\frac{{\rm C}^{\rm{plane}}_{I}}{\kappa\epsilon_0\, \epsilon_m}\right)e^{-\kappa a}\right]^{2}}\right\}\,,
\label{eq:planepressureap}
\end{eqnarray}
\end{widetext} 
which enables us to finally calculate the interaction potential energy per unit area \cite{butt10}
\begin{equation}
u_{\rm{planes}}(L)=-\int_{\infty}^{L}P\left(a\right)da\,,
\label{eq:uplanesap}
\end{equation}
and the force in the PFA regime by \cite{butt10,israelachvili11}
\begin{eqnarray}
\rm{F}_{\rm D}^{\rm (PFA)}(\mathit{L})&=&2\pi\left(\frac{R_{1}R_{2}}{R_{1}+R_{2}}\right)u_{\rm{planes}}(L)\,.
\label{eq:fPFAap}
\end{eqnarray}
Finally, using (\ref{eq:planepressureap}) in (\ref{eq:uplanesap}) and the result in (\ref{eq:fPFAap}) leads us to a quadratic function of $\psi_M$, where the corresponding coefficients
 $\mathcal{A}_{\rm D}^{\rm (PFA)}$,  $\mathcal{B}_{\rm D}^{\rm (PFA)}$, and  $\mathcal{C}_{\rm D}^{\rm (PFA)}$ are given by
\begin{widetext}
\begin{eqnarray}
\mathcal{A}_{\rm D}^{\rm (PFA)}\left(L\right)&=&-4\pi\epsilon_0\, \epsilon_m\kappa^{2}\left(\frac{R_{1}R_{2}}{R_{1}+R_{2}}\right)\left[1-\left(\frac{{\rm C}^{\rm{plane}}_{I}}{\kappa\epsilon_0\, \epsilon_m}\right)^{2}\right]\int_{\infty}^{L}\frac{1}{\left[\left(1+\frac{{\rm C}^{\rm{plane}}_{I}}{\kappa\epsilon_0\, \epsilon_m}\right)e^{\kappa a}+\left(1-\frac{{\rm C}^{\rm{plane}}_{I}}{\kappa\epsilon_0\, \epsilon_m}\right)e^{-\kappa a}\right]^{2}}da\,,\nonumber\\
\mathcal{B}_{\rm D}^{\rm (PFA)}\left(L\right)&=&-4\pi\kappa\bar{\sigma}_{1}\left(\frac{R_{1}R_{2}}{R_{1}+R_{2}}\right)\int_{\infty}^{L}\frac{\left[\left(1+\frac{{\rm C}^{\rm{plane}}_{I}}{\kappa\epsilon_0\, \epsilon_m}\right)e^{\kappa a}-\left(1-\frac{{\rm C}^{\rm{plane}}_{I}}{\kappa\epsilon_0\, \epsilon_m}\right)e^{-\kappa a}\right]}{\left[\left(1+\frac{{\rm C}^{\rm{plane}}_{I}}{\kappa\epsilon_0\, \epsilon_m}\right)e^{\kappa a}+\left(1-\frac{{\rm C}^{\rm{plane}}_{I}}{\kappa\epsilon_0\, \epsilon_m}\right)e^{-\kappa a}\right]^{2}}da\,,\nonumber\\
\mathcal{C}_{\rm D}^{\rm (PFA)}\left(L\right)&=&\frac{4\pi\bar{\sigma}_{1}^{2}}{\epsilon_0\, \epsilon_m}\left(\frac{R_{1}R_{2}}{R_{1}+R_{2}}\right)\int_{\infty}^{L}\frac{1}{\left[\left(1+\frac{{\rm C}^{\rm{plane}}_{I}}{\kappa\epsilon_0\, \epsilon_m}\right)e^{\kappa a}+\left(1-\frac{{\rm C}^{\rm{plane}}_{I}}{\kappa\epsilon_0\, \epsilon_m}\right)e^{-\kappa a}\right]^{2}}da.
\end{eqnarray}
\end{widetext}
\par To calculate the force in the PFA approximation between one metallic microsphere at a fixed electrostatic potential $\psi_{\rm{M}}$ and an isolated metallic microsphere with fixed charge $Q$, both immersed in an isotropic $Z:Z$ electrolyte with relative permittivity $\epsilon_{m}$ and in thermal equilibrium at a temperature $T$, we have to consider the boundary value problem of two half regions $I$ and $I\!I$, being one held at $\psi_{\rm{M}}$ with surface at $z=0$ and the other electrically isolated with a constant surface charge density $\sigma_{1}=Q/A$ and surface at $z=a$. The corresponding equations are still given by (\ref{eq:PFAequations}), but now the boundary conditions are
\begin{eqnarray}
\left.\psi^{I\!I}(z)\right|_{z=0^{-}}&=&\left.\psi^{I\!I\!I}(z)\right|_{z=0^{+}}\,,\label{eq:bcPFA1aIso}\\
-\epsilon_{m}\left.\frac{d\psi^{I\!I\!I}}{dz}(z)\right|_{z=0^{+}}&=&\frac{\sigma_{2}(\kappa a)}{\epsilon_{0}}\,,\label{eq:bcPFA1bIso}\\
\left.\psi^{I\!I\!I}(z)\right|_{z=a^{-}}&=&\left.\psi^{I}(z)\right|_{z=a^{+}}\,,\label{eq:bcPFA2aIso}\\
\epsilon_{m}\left.\frac{d\psi^{I\!I\!I}}{dz}(z)\right|_{z=a^{-}}&=&\frac{\sigma_{1}}{\epsilon_{0}}\,,\label{eq:bcPFA2bIso}
\end{eqnarray}
Using (\ref{eq:planesol}) on (\ref{eq:bcPFA1aIso}) and (\ref{eq:bcPFA2bIso}), we then get
\begin{eqnarray}
\mbox{A}&=&\frac{\frac{\sigma_{1}}{\kappa\epsilon_0\, \epsilon_m}+e^{-\kappa a}\psi_{\rm{M}}}{e^{\kappa a}+e^{-\kappa a}}\,,\label{eq:planeAIso}\\
\mbox{B}&=&\frac{-\frac{\sigma_{1}}{\kappa\epsilon_0\, \epsilon_m}+e^{\kappa a}\psi_{\rm{M}}}{e^{\kappa a}+e^{-\kappa a}}.
\label{eq:planeBIso}
\end{eqnarray}
As a result, (\ref{eq:pressurePFA}) is given by
\begin{eqnarray}
P(a) &=& \frac{2}{\epsilon_0\, \epsilon_m}\times\nonumber\\
&&\left[\frac{\left(\kappa\epsilon_0\, \epsilon_m\right)^{2}\psi^{2}_{M}+\kappa\epsilon_0\, \epsilon_m\sigma_{1}\left(e^{\kappa a}-e^{-\kappa a}\right)\psi_{\rm{M}}-\sigma_{1}^{2}}{\left(e^{\kappa a}+e^{-\kappa a}\right)^{2}}\right]\,,\nonumber\\
&&\label{eq:planepressureapIso}
\end{eqnarray}
which allows us to calculate the interaction potential energy per unit area (\ref{eq:uplanesap}),
\begin{equation}
u_{\rm{planes}}(L)=\frac{\kappa\epsilon_0\, \epsilon_m e^{-\kappa L}\psi^{2}_{M}+2\sigma_{1}\psi_{\rm{M}}-\frac{\sigma^{2}_{1}}{\kappa\epsilon_0\, \epsilon_m}e^{
-\kappa L}}{e^{\kappa L}+e^{-\kappa L}}\,,
\label{eq:uplanesapIso}
\end{equation}
and the force in the PFA regime
\begin{eqnarray}
\rm{F}_{\rm M}^{\rm (PFA)}(\mathit{L})&=&2\pi\left(\frac{R_{1}R_{2}}{R_{1}+R_{2}}\right)\nonumber\\
&&\times\frac{\kappa\epsilon_0\, \epsilon_m e^{-\kappa L}\psi^{2}_{\rm{M}}+2\sigma_{1}\psi_{\rm{M}}-\frac{\sigma^{2}_{1}}{\kappa\epsilon_0\, \epsilon_m}e^{
-\kappa L}}{e^{\kappa L}+e^{-\kappa L}}\,.\nonumber\\
&&\label{eq:fPFAapIso}
\end{eqnarray}
%


\bibliography{Casimir-Dlayer-met}

\end{document}